\documentclass[%
reprint,
superscriptaddress,
amsmath,amssymb,
aps,
prb,
floatfix,
]{revtex4-2}

\usepackage{graphicx}
\usepackage{dcolumn}
\usepackage{bm}
\usepackage{xcolor}
\usepackage{xr-hyper}
\usepackage{xr}
\usepackage{hyperref}
\usepackage{amsmath}
\usepackage{dirtytalk}

\begin{document}

\preprint{APS/123-QED}

\title[PRB]{Lifetime effects and satellites in the photoelectron spectrum of tungsten metal}

\author{C.~Kalha}
\affiliation{Department of Chemistry, University College London, 20 Gordon Street, London, WC1H~0AJ, United Kingdom.}

\author{L.~E.~Ratcliff}
\affiliation{Department of Materials, Imperial College London, London, SW7 2AZ, United Kingdom}
\affiliation{Thomas Young Centre for Theory and Simulation of Materials, London, UK}

\author{J.~J.~Gutiérrez Moreno}
\affiliation{Barcelona Supercomputing Center (BSC), C/ Jordi Girona 31, 08034 Barcelona, Spain.}

\author{S.~Mohr}
\affiliation{Barcelona Supercomputing Center (BSC), C/ Jordi Girona 31, 08034 Barcelona, Spain.}
\affiliation{Nextmol (Bytelab Solutions SL), C/ Roc Boronat 117, 08018 Barcelona, Spain.}

\author{M.~Mantsinen}
\affiliation{Barcelona Supercomputing Center (BSC), C/ Jordi Girona 31, 08034 Barcelona, Spain.}
\affiliation{ICREA, Pg. Lluís Companys 23, 08034 Barcelona, Spain.}

\author{N.~K.~Fernando}
\affiliation{Department of Chemistry, University College London, 20 Gordon Street, London, WC1H~0AJ, United Kingdom.}

\author{P.~K.~Thakur}
\author{T.-L.~Lee}
\affiliation{Diamond Light Source Ltd., Harwell Science and Innovation Campus, Didcot, Oxfordshire, OX1 3QR, United Kingdom.}

\author{H.-H.~Tseng}
\author{T.~S.~Nunney}
\affiliation{Thermo Fisher Scientific, Surface Analysis, Unit 24, The Birches Industrial Estate, East Grinstead, West Sussex, RH19 1UB, United Kingdom.}

\author{J.~M.~Kahk}
\affiliation{Department of Materials, Imperial College London, London, SW7 2AZ, United Kingdom}
\affiliation{Thomas Young Centre for Theory and Simulation of Materials, London, UK}
\affiliation{Institute of Physics, University of Tartu, W. Ostwaldi 1, 50411 Tartu, Estonia}

\author{J.~Lischner}
\affiliation{Department of Materials, Imperial College London, London, SW7 2AZ, United Kingdom}
\affiliation{Thomas Young Centre for Theory and Simulation of Materials, London, UK}

\author{A.~Regoutz}
 \email{a.regoutz@ucl.ac.uk}
\affiliation{Department of Chemistry, University College London, 20 Gordon Street, London, WC1H~0AJ, United Kingdom.}

\date{\today}

\begin{abstract}

Tungsten is an important and versatile transition metal and has a firm place at the heart of many technologies. A popular experimental technique for the characterisation of tungsten and tungsten-based compounds is X-ray photoelectron spectroscopy (XPS), which enables the assessment of chemical states and electronic structure through the collection of core level and valence band spectra. However, in the case of metallic tungsten, open questions remain regarding the origin, nature, and position of satellite features that are prominent in the photoelectron spectrum. These satellites are a fingerprint of the electronic structure of the material and have not been thoroughly investigated, at times leading to their misinterpretation. The present work combines high-resolution soft and hard X-ray photoelectron spectroscopy (SXPS and HAXPES) with reflection electron energy loss spectroscopy (REELS) and a multi-tiered ab-initio theoretical approach, including density functional theory (DFT) and many-body perturbation theory (G0W0 and GW+C), to disentangle the complex set of experimentally observed satellite features attributed to the generation of plasmons and interband transitions. This combined experiment-theory strategy is able to uncover previously undocumented satellite features, improving our understanding of their direct relationship to tungsten's electronic structure. Furthermore, it lays the groundwork for future studies into tungsten based mixed-metal systems and holds promise for the re-assessment of the photoelectron spectra of other transition and post-transition metals, where similar questions regarding satellite features remain.

\end{abstract}

\maketitle

\section{Introduction}

Interest in tungsten (W) metal is seeing a resurgence owing to its high sputtering threshold,~\cite{Smid1998DevelopmentComponents} low hydrogen retention,~\cite{Fujita2016EffectTungsten, Causey2001TheProperties} and high temperature resistance,~\cite{Causey1999TritiumTritons} all of which make it extremely attractive as a plasma facing material in nuclear fusion reactors.~\cite{Abernethy2017PredictingReview, Philipps2011TungstenDevices} Additionally, tungsten forms the base of many technologies in a range of industrial fields, with tungsten oxides being widely used for their electronic, photoabsorption, optical, and catalytic properties.~\cite{Niklasson2007ElectrochromicsThese, Walter2010SolarCells, Migas2010TungstenPhases, Zheng2011NanostructuredApplications, Mardare2019ReviewCoatings} Tungsten-based alloys and intermetallic compounds find applications as diffusion barriers in metallisation schemes for semiconductor devices due to their chemical inertness towards the surrounding materials and high chemical blocking efficiency.~\cite{Plappert2012CharacterizationSpectroscopy, Fugger2014ComparisonTi, Roshanghias2014OnCircuits, Kleinbichler2017AnnealingLayers} The functionality and reliability of these materials is to a large extent governed by the electronic structure of the base metal.~\cite{Wach2020ComparativeSpectroscopies} Moreover, when characterising the properties of tungsten-based compounds, studies often necessitate comparison to those of tungsten metal to elucidate differences. Therefore, the investigation and accurate determination of the electronic structure of tungsten metal is highly relevant today and critical for the design of new materials and implementation of new tungsten-based technologies.\par

The characterisation of the electronic structure of metallic tungsten using both theoretical and experimental approaches has a long-standing history spanning many decades. The first calculations of the electronic band structure of tungsten were reported by Manning~\textit{et al.}\ in 1939.~\cite{Manning_1939} Studies have followed continuously since then.~\cite{Mattheiss1965FermiTungsten, Loucks1965FermiMethod, Petroff1971CalculationMolybdenum, Feder1975Spin-orbitMetals, Posternak1980Self-consistentW001, Mattheiss1984ElectronicSurface, Jansen1984Total-energyTungsten, Wei1985LinearizedTungsten, Legoas2000Self-consistentSurfaces} Most notably, Mattheiss~\textit{et al.}~\cite{Mattheiss1965FermiTungsten}\ provided the first theoretical study investigating the Fermi surface of tungsten using a non-relativistic approach, followed by Christensen~\textit{et al.}, who expanded on this study by calculating the band structure using a relativistic augmented-plane-wave (APW) method coupling the theory results with experimentally obtained photoelectron spectra.~\cite{Feuer_I, Feuer_II} More recently, theoretical investigations have transitioned from calculating tungsten's band structure to more application-driven investigations into the interaction of molecules with tungsten surfaces, point defect studies, or the study of nanostructures.~\cite{Jiang2010StrongCalculations, Ventelon2012AbMetals, Sun2014ElectronicCalculation, Fernandez2015HydrogenModels, Zhang2017TrappingStudy, Song2018StructureStudy} Several approaches have been used to calculate the projected density of states (PDOS) of tungsten, but these results have not yet been directly compared (with photoionisation cross section and broadening corrections) to high-resolution valence band spectra.~\cite{Feuer_I, Petroff1971CalculationMolybdenum, Posternak1980Self-consistentW001, Legoas2000Self-consistentSurfaces, Wach2020ComparativeSpectroscopies} Beyond band structure features, the photoelectron spectrum also contains satellite peaks which arise from the additional excitation of a plasmon or electron-hole pair. To date, no theoretical study of satellites in tungsten has been reported. Such a treatment requires the use of many-body perturbation theory (MBPT), which has already shown promise for the determination of plasmon satellites in sodium and aluminium,~\cite{Aryasetiawan1996MultipleExpansion}, and silicon,~\cite{Guzzo_2012, Lischner2013PhysicalStudy, Lischner2015SatelliteCoupling}, but also heavier transition metal compounds.~\cite{Jiang_2012, Regoutz2019InsightsTheory, Lu2021Layer-resolvedSpectroscopy} 

Alongside these theoretical studies, several groups have experimentally explored the electronic structure, interfacial properties, and surface and bulk effects of tungsten, using a range of techniques, including soft angle-resolved photoelectron spectroscopy (ARPES),~\cite{Hussain1980Temperature-dependentEffects, Jugnet1987AngleW110, Gaylord1987Spin-orbit-interaction-inducedW011}, hard X-ray ARPES (HARPES)~\cite{Gray2011ProbingPhotoemission}, soft and hard X-ray photoelectron momentum microscopy~\cite{Medjanik_2017, Medjanik2019ProgressRecording}, inverse photoemission spectroscopy (IPES),~\cite{Drube1986UnoccupiedPhotoemission} X-ray photoelectron spectroscopy (XPS),~\cite{Penchina1974PhotoemissionFilms, Chen1995AdsorptionW100, Warren1996OxidationAtmospheres, Engelhard2000ThirdSpectroscopy, Colton1976ElectronicSpectroscopy, Egawa1983ADSORPTIONSTUDIES} ultra-violet photoelectron spectroscopy (UPS),~\cite{Feuerbacher1973DirectionalFaces, Feuer_II, Penchina1974PhotoemissionFilms, Feydt1998PhotoemissionW110, VanDerVeen1982Chemisorption-inducedTa111, Egawa1983ADSORPTIONSTUDIES} and synchrotron-based photoelectron spectroscopy.~\cite{Smith1976StudyRadiation, Holmes1981DispersionReconstruction, Mueller1988AOtextsubscript2, Riffe1990Core-holeW110, Mullins1993Sulfur-inducedShift} \par

Despite the extensive existing body of both theoretical and experimental work on tungsten metal, there are still aspects of tungsten's electronic structure and its influence on photoelectron spectra that have not been fully explored and understood. One particular aspect that warrants further investigation is the presence of plasmon satellites. The photoexcitation of electrons in metallic-like systems generates a final state effect, known as plasmon satellites, which appear as features on the higher binding energy (BE) side of the main ionisation peaks in photoelectron spectroscopy experiments, and originate from the coupling of the core hole and interaction of photoelectrons with conduction electrons. Such satellites present many challenges when analysing spectra and are rarely considered or in some cases misinterpreted as additional chemical states. Several experimental studies have explored plasmon satellites in the photoelectron spectra of metals, but these have been limited to first-row or noble ``simple'' metals,~\cite{Ley1975Many-bodyMagnesium, SimpleMetals_1978, Th1979Bulk-andMetal, Bates_1979, Kurth2003DeterminationMagnesium, Leiro1983StudyGold} with a lack of investigation into the satellite structure of heavier transition metals, such as tungsten. Plasmon satellites have been confirmed in past electron energy loss spectroscopy studies on tungsten~\cite{Weaver1975OpticalTungsten, Luscher1977EnergySpectroscopy}, but to the best of our knowledge no XPS study has been reported that captures or discusses these satellites and their influence on the photoelectron spectrum.\par

Besides the presence of plasmon satellites, the core level peaks of tungsten recorded by photoelectron spectroscopy have their own inherent challenges. The two most frequently accessed core levels are the shallow W~4\textit{f} (31-34~eV) and W~4\textit{d} (240-260~eV).~\cite{Engelhard2000ThirdSpectroscopy} The 4\textit{f} spectrum is particularly difficult to analyse as the 4\textit{f} peaks possess a narrow full-width at half maximum (FWHM), but also the 5\textit{p}\textsubscript{3/2} core line lies in close proximity to the 4\textit{f}\textsubscript{5/2}
line, and so must be considered and included, if a peak-fit analysis is required. Additionally, if both the metal and tetravalent (IV) oxidation states are present, the 5\textit{p}\textsubscript{3/2} metal core line overlaps with the 4\textit{f} doublet peak of the W(IV) state.~\cite{Kalha2021ThermalSpectroscopy} The W~4\textit{d} core level exhibits a large lifetime broadening, leading to a significant Lorentzian contribution to the line shape, which is often difficult to describe when peak fitting. Therefore, the presence of satellites and their lack of characterisation in tungsten photoelectron spectra, coupled with the complexity of the shallow core levels, form a strong motivation to revisit the photoelectron spectrum and electronic structure of tungsten.\par

The present work combines soft and hard X-ray photoelectron spectroscopy (SXPS and HAXPES) to study the satellite structure of key tungsten core levels, as well as providing high-resolution valence band spectra. HAXPES enables the exploration of bulk tungsten by minimising the pure surface nature of specific spectral features. In addition, it allows access to deeper core levels, which add complementary information to the common core states studied with SXPS and may offer a solution to the challenges associated with the interpretation of the complex W~4\textit{f} and 4\textit{d} core levels. Reflection electron energy loss spectroscopy (REELS) is used in parallel to directly determine the energy loss features of tungsten and to aid in the assignment of satellite features observed in photoelectron spectroscopy.\par

Given the complexity of the experimental spectra of tungsten, theoretical modelling is required to aid the interpretation of the spectral features, and this forms the primary motivation for the re-calculation of the electronic structure of tungsten. In order to fully analyse and interpret the complex electronic structure of tungsten, experiments are complemented with a multi-tiered theory approach. Density functional theory (DFT)~\cite{Hohenberg_1964, Kohn_1965} is combined with MBPT within the GW and ``GW plus cumulant'' (GW+C) approaches.~\cite{Hybertsen1986ElectronEnergies, Guzzo2011ValenceSatellites, Lischner2013PhysicalStudy, Caruso2015BandPolarons} This allows the identification of specific observed spectral features arising from the electronic structure of tungsten, including the various satellite features. Additionally, given the interest in tungsten-based alloys in the semiconductor industry, linear-scaling DFT (LS-DFT) is used and compared to conventional cubic-scaling DFT. LS-DFT is able to model many-thousand atom systems by overcoming the computational cost limitations of cubic-scaling DFT, and thus is useful for the accurate description of disordered mixed-metal alloys in device systems in future studies.~\cite{Mohr2014DaubechiesTheory,Mohr2015AccurateApplicability}\par

\section{Experimental Methodology}

A polycrystalline tungsten foil (99.95~at.\% metal basis, 0.1~mm thick, Goodfellow Cambridge Ltd.) was used for the REELS, SXPS, and HAXPES measurements. Details regarding the ex- and in-situ preparation of the sample for the different measurements can be found in the Supplementary Information~I. REELS measurements were conducted on a Thermo Scientific Nexsa XPS instrument, employing its flood gun as the electron source, with a beam energy of 1~keV and emission current of 5~$\mu$A. Back scattered electrons were measured using a 180$^{\circ}$ hemispherical analyser in conjunction with a one-dimensional detector. A pass energy of 40~eV was used to collect the REELS data.\par

SXPS measurements were conducted on a Thermo K-Alpha XPS instrument, which operates with a monochromatised Al~K$\alpha$ excitation source (1.4867~keV) and consists of a hemispherical analyser and two-dimensional detector. Measurements were conducted with a 400~$\mu$m elliptical spot size, 6~$\mu$A X-ray anode emission current, 30~$\mu$A flood gun emission current, and at a base pressure of 2$\times$10\textsuperscript{-9}~mbar. Survey, core level, and valence band spectra were collected with a pass energy of 200~eV, 20~eV, and 15~eV, respectively. HAXPES measurements were conducted on beamline I09 at the Diamond Light Source, UK.~\cite{Lee2018ASource} A photon energy of 5.9267~keV (further referred to as 5.93~keV for simplicity) was selected using a double crystal Si~(111) monochromator and an additional Si~(004) channel-cut postmonochromator. The end station is equipped with a VG Scienta EW4000 electron analyser with a $\pm$28$^{\circ}$ angular acceptance. All measurements were performed in grazing incidence geometry at angles below 5$^\circ$ between the incoming X-ray beam and the sample surface. A pass energy of 200~eV was used for the collection of all spectra. The experimental resolution was evaluated by measuring the intrinsic Fermi edge of the tungsten sample and fitting the data with a Gaussian-broadened Fermi-Dirac distribution. From this the resolution of the SXPS and HAXPES measurements were determined to be 350 and 266~meV, respectively (see Supplementary Information~II). The probing depth of the SXPS and HAXPES measurements is approximately three times the inelastic mean free path (IMFP) of the photoelectrons. The total IMFP of all electrons in tungsten was calculated using the TPP-2M formula resulting in values of 2.02 and 6.01~nm for the soft and hard X-ray excitation energies, respectively.~\cite{Tanuma1993CalculationsRange}

\section{Computational Methodology}
\subsection{Density Functional Theory}\label{DFT}

When comparing DFT with valence XPS, a theoretical spectrum can be generated by calculating a partial density of states (PDOS), and applying appropriate photoionisation cross sections, as discussed below. The accuracy of the resulting spectrum thus depends on both the accuracy of the calculated energy bands, which is influenced by factors including the choice of exchange-correlation functional, basis set convergence, use of pseudopotentials and level of $k$-point sampling, and the details of the projection, i.e.\ choice of atomic orbitals and projection scheme. Since there is no unambiguous choice of either projection scheme or localized atomic basis, it is therefore important to consider the influence of different approaches. To this end, DFT calculations were performed with two different codes, the plane-wave Quantum Espresso code~\cite{Giannozzi2009QUANTUMMaterials} and the wavelet-based BigDFT code~\cite{Ratcliff2020FlexibilitiesCalculations}, which use different basis sets, pseudopototentials, and projection schemes. For both sets of calculations the Perdew-Burke-Ernzerhof (PBE) exchange-correlation functional was employed.~\cite{Perdew1996GeneralizedSimple} For simplicity, the two approaches will be referred to by the basis set used for the calculations when discussing the results.

Quantum Espresso calculations were performed in the primitive unit cell, with a lattice parameter of 3.184~\AA, with a 16~$\times$~16$\times$~16 \textit{k}-point grid.  In order to obtain a smooth density of states, the DFT eigenvalues were then interpolated onto a 64~$\times$~64$\times$~64 \textit{k}-point grid. The PDOS was generated using a L\"{o}wdin population analysis-based approach.~\cite{Lowdin1950OnCrystals, Lowdin1970OnProblem} Further computational details, including the local orbitals used to perform the projection, are given in Supplementary Information~III.\par

BigDFT calculations were performed using the linear scaling version of the code,~\cite{Mohr2014DaubechiesTheory,Mohr2015AccurateApplicability}, since the localized and in-situ optimised atom-centred support function basis provides a natural and accurate approach for generating the PDOS, using a Mulliken-type projector~\cite{Mulliken1955ElectronicI} onto the support functions.~\cite{Mohr2017ComplexityBasis, Dawson2020ComplexityEmbedding}. Furthermore, as previously discussed, the use of LS-DFT will be valuable for future studies of disordered mixed-metal systems, and it is thus important to compare LS-DFT results with experiment for bulk tungsten. As the first Brillouin Zone (BZ) is only sampled at the $\Gamma$-point in linear scaling BigDFT, $k$-point sampling is not available and so calculations were instead performed in a 12~$\times$~12~$\times$~12 body centered cubic (BCC) supercell with a side dimension of 38.804~{\AA} (a = 3.234~{\AA}), comprising 3456 atoms. For such a large supercell, $\Gamma$-point sampling was shown to be sufficient to reach total energy convergence. Additionally, 1458, 2000, and 2662 atom models were simulated (see Supplementary Information~IV) to assess the supercell convergence. Further computational details are given in Supplementary Information~III.


\subsection{G0W0, and GW+C}\label{GW,GWC}

Full frequency G0W0 calculations were carried out in BerkeleyGW~\cite{Deslippe2012BerkeleyGW:Nanostructures} using the PBE eigenstates calculated using Quantum Espresso. The frequency-dependent dielectric matrix was calculated within the random phase approximation. Next, the frequency-dependent electronic self-energy was calculated for all eigenstates at all \textit{k}-points in the symmetry-reduced \textit{k}-point grid. We included the static remainder correction in the Coulomb hole term as described in Ref.~\cite{Deslippe2013Coulomb-holeApproach}. Using the frequency dependent self energies, GW+C spectral functions were computed as described in Ref.~\cite{Lischner2015SatelliteCoupling}. As with the plane-wave DFT eigenvalues, the G0W0 eigenvalues and the GW+C spectral functions were interpolated onto a 64$\times$64$\times$64 \textit{k}-point grid. Full details are provided in the Supplementary Information III.

The GW+C spectral functions from the BerkeleyGW calculations were also used to model the core level photoelectron spectra. The W~4\textit{f} and W~5\textit{p} electrons were explicitly included in the calculations. The calculated 4\textit{f} and 5\textit{p} spectral functions (at the $\Gamma$ point) were used to reconstruct the experimental 4\textit{f}/5\textit{p} core level spectrum, as follows: for each subshell, two copies of the calculated spectral function were added together, with one copy shifted by the atomic spin orbit splitting, determined from HAXPES measurements and weighted by the theoretical intensity ratio for the spin orbit doublet determined from the tabulated Scofield photoionisation cross section tabulated data~\cite{Scofield1973TheoreticalKeV}. The doublet peaks were then broadened to reflect the intrinsic lifetime broadening due to radiative recombination and Auger decay, as well as the experimental broadening. The resultant 4\textit{f} and 5\textit{p} simulated doublets were then scaled relative to each other using the Scofield cross sections and shifted accordingly to match the appearance of the experimental spectrum. The deeper core levels were not explicitly included in the calculations (they were contained within the pseudopotential). However, we note that the satellite structures of the different core levels are similar to each other (Fig.~\ref{fig:Satellite_Comparison}). Based on this observation, we have chosen to also use the calculated 4\textit{f} spectral function to construct theoretical 3\textit{d} and 4\textit{d} spectra, using the same method as described above. We have recently used a similar approach for predicting core level line shapes in PdCoO\textsubscript{2}.~\cite{Lu2021Layer-resolvedSpectroscopy} A detailed explanation along with the values used to construct these simulated core level spectra can be found in the Supplementary Information~V.

\subsection{Comparison of Theory and Experiment}\label{X_section_method}

To provide a direct comparison between the theoretically calculated PDOS and the experimental valence band (VB) spectra, the PDOS was aligned to the calculated Fermi energy ($E_F$) from the respective calculations, and the VB spectra were aligned to the experimentally observed Fermi edge. Furthermore, the individual PDOS contributions require weighting according to their respective photoionisation cross sections ($\sigma_i$) at the photon energies used in the experiments. However, theoretical cross sections are only available for states which are occupied in the ground state of the atom. This presents a limitation in the case of tungsten, where the contribution from the \textit{p} states is significant, and originates from mixing of the unoccupied 6\textit{p} conduction band state. The 5\textit{p} orbital is a shallow core level, at 38~eV above the Fermi energy, and is unlikely to contribute to states within the VB. Nevertheless, applying the theoretical cross sections for the 5\textit{p}, 6\textit{s} and 5\textit{d} orbitals provides good agreement to experiment (see Supplementary Information~VI). However, this approach is not well justified.\par

Mudd~\textit{et al.}~\cite{Mudd_2014} encountered a similar challenge, in the case of CdO, where the unoccupied Cd~5\textit{p} state contributes to the valence \textit{p} character. The Cd~5\textit{p} orbital, much like the W~6\textit{p} orbital, is unoccupied in the ground state of the atom, and so Mudd~\textit{et al.}\ approached the problem by multiplying the In~5\textit{p}/In~5\textit{s} cross section ratio to the cross section of the Cd~5\textit{s} orbital to estimate the Cd~5\textit{p} cross section. Indium was chosen as it is the first element to have an electron in the 5\textit{p} orbital. Using this approach for tungsten, the ratio of Pb~6\textit{p}/Pb~6\textit{s} was multiplied with the W~6\textit{s} cross section (values taken from Refs.~\cite{Scofield1973TheoreticalKeV, Kalha20}, but the resulting 6\textit{p} cross section had almost negligible contribution to the simulated photoelectron spectrum (i.e.\ the sum of the weighted density of states), and did not result in a good agreement between experiment and theory (see Supplementary Information~VI).\par

Another example is the case of metallic silver, where Panaccione~\textit{et al.}~\cite{Panaccione_2005}\ attributed the ``free electron-like character'' of the unoccupied Ag~5\textit{p} orbital to the valence \textit{p} orbital character. In contrast to the method used by Mudd~\textit{et al.}, Panaccione~\textit{et al.}\ applied a fitting procedure to optimise the weight of \textit{s}, \textit{p} and \textit{d} contributions and therefore indirectly determine the 5\textit{p} cross section. To apply this method to tungsten, the weighting factors of the \textit{s} and \textit{d} states were first constrained to the Scofield tabulated cross section values of the 6\textit{s} and 5\textit{d} states (see Supplementary Information~VI) at the given excitation energy.~\cite{Scofield1973TheoreticalKeV} Then the \textit{p} state weighting was determined by minimising the sum of the least squared difference between the simulated spectrum and experimental spectrum. This approach gave much better agreement (see Supplementary Information~VI) to the experimental spectrum.\par

The comparison between the three approaches applied to the plane-wave DFT PDOS -- (1) using the W~5\textit{p} cross section (implemented using the Galore software package~\cite{JJackson2018Galore:Spectroscopy}), (2) determining the 6\textit{p} cross section using cross sections of Pb, and (3) the ``optimised'' method outlined by Panaccione~\textit{et al.}\ are displayed in the Supplementary Information~VI, with the ``optimised'' approach providing the most suitable weighting. When comparing theory to experiment in Section~\ref{sec:Electronic}, the ``optimised'' W~6\textit{p} state cross section (determined from optimising the G0W0 PDOS) and the Scofield W~6\textit{s} and W~5\textit{d} cross sections were used.\par

\section{Results and Discussion}

\subsection{REELS}

\begin{figure}[ht!]
\centering
    \includegraphics[keepaspectratio, width = 0.45\textwidth]{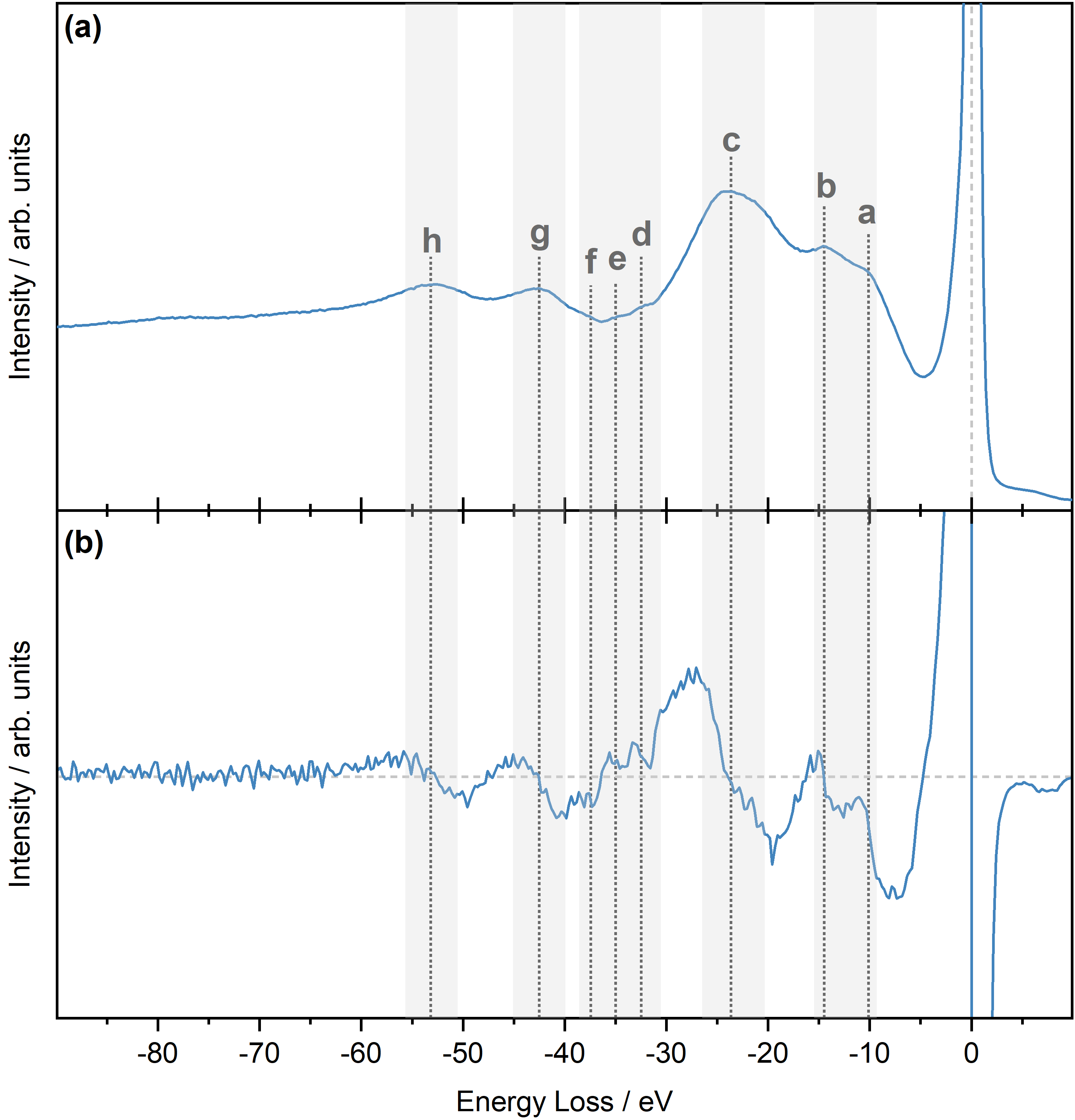}
    \caption{REELS spectrum of metallic tungsten. (a) Raw REELS spectrum, and (b) the first derivative of the REELS spectrum. Features of interest are labelled as \textbf{a-h}. The spectra are aligned so that the primary elastic peak is at 0~eV. The dotted horizontal line in (b) indicates y~=~0.}
    \label{fig:REELS}
\end{figure}

A number of studies have used electron energy loss spectroscopies or optical techniques to probe the electronic excitations in tungsten.~\cite{Harrower1956AugerW, Tharp1967EnergyTungsten, Porteus1970ExcitationDiffraction, Edwards1971EnergyTungsten, Burkstrand_1972, Stein1972PlasmonCrystals, Weaver1975OpticalTungsten, Luscher1977EnergySpectroscopy,Avery_1981, Gergely1981ElasticSpectroscopy, Shinar1984TheTungsten, Czy_1990, Czy_1991} These measurements provide a basis to help identify satellite features in the SXPS and HAXPES core level spectra, as will be discussed in Section~\ref{XPS}, motivating the collection of a high-resolution REELS spectrum using a bulk-sensitive incident electron energy (see Fig.~\ref{fig:REELS}(a)). The first derivative of the energy loss intensity is displayed in Fig.~\ref{fig:REELS}(b) to aid with the identification of energy loss peaks and their energy positions. Prominent peaks are identified within five regions located between 10-54~eV, and labelled with letters \textbf{(a-h)} in Fig.~\ref{fig:REELS}(a). Table~\ref{REELS_peak} lists the energy loss ($w$) positions of all identifiable peaks.\par

\begin{table}[ht!]
    \caption{\label{REELS_peak}Energy loss regions and peak positions extracted from the recorded REELS spectrum as well as the feature notation used throughout.}
    \begin{ruledtabular}
    \begin{tabular}{ccc}
    Energy Loss Region / eV & Feature & \textit{w}~/~eV  \\
    \hline
    10-15 & a & 10.1  \\
    & b & 14.5  \\
    21.8-25.5 & c & 23.7  \\
    30.5-39 & d & 32.5  \\
    & e & 35.0  \\
    & f & 37.5  \\
    40.0-45.5 & g & 42.5 \\
    50.5-55.8 & h & 53.2  \\
    \end{tabular}
    \end{ruledtabular}
\end{table}

Since the first reported electron energy loss measurements on tungsten by Harrower,~\cite{Harrower1956AugerW} the origin of the observed loss peaks has been subject to continuous discussion but a definitive interpretation of the observed features remains outstanding. A good starting point for the interpretation of REELS data is to first determine the theoretical values of the surface and bulk plasmons using the Langmuir equation derived for a homogeneous electron gas.~\cite{Ritchie57, SurfSci2003} The theoretical bulk and surface plasmon energies of tungsten are estimated to be 22.8 and 16.2~eV, respectively, assuming six valence electrons per atom (5\textit{d}\textsuperscript{4}6\textit{s}\textsuperscript{2}).\par

The most intense feature \textbf{c} in Fig.~\ref{fig:REELS} is located at an energy loss of 23.7~eV from the primary elastic peak at 0~eV. This is assigned as the bulk plasmon and is in good agreement with the theoretical bulk plasmon energy and values obtained in previous studies.~\cite{Weaver1975OpticalTungsten, Luscher1977EnergySpectroscopy, Avery_1981} According to Weaver~\textit{et al.}\ the reason for the slight shift from the theoretical value is due to the existence of interband transitions close to the 25~eV region.~\cite{Weaver1975OpticalTungsten}\par

Weaver,~\cite{Weaver1975OpticalTungsten} Luscher,~\cite{Luscher1977EnergySpectroscopy} and Avery~\cite{Avery_1981} all observe that the surface plasmon of tungsten is much higher in energy than 16.2~eV, instead they assign a peak at approximately 20-21~eV to the surface plasmon. A shift from the theoretical value was also observed by Weaver~\textit{et al.}\ for other body centred cubic (BCC) transition metals (Nb, V, Ta, Mo), who attributed this observation to screening effects.~\cite{Weaver1973OpticalEV, Weaver1974OpticalEV} The surface plasmon is difficult to observe in Fig.~\ref{fig:REELS} as the incident electron energy is considerably higher than those of past studies, and therefore the collected data is dominated by the bulk plasmon. The region in which peak \textbf{c} resides does appear slightly asymmetric on the lower energy loss side and a secondary peak, perhaps the surface plasmon, may be present in this region. Avery also reported difficulty in resolving features in this energy loss region of tungsten using a 901~eV excitation energy.~\cite{Avery_1981}\par

Two additional features, \textbf{a} and \textbf{b} are identified in the low-energy loss region at 10.1~eV and 14.5~eV, respectively. Several studies report peaks in this region for tungsten~\cite{Luscher1977EnergySpectroscopy, Weaver1975OpticalTungsten, Avery_1981, Shinar1984TheTungsten}, which are also found in other BCC transition metals~\cite{Weaver1974OpticalEV, Romaniello2006OpticalEV}. Shinar~\textit{et al.}\ summarises the discussion around the exact nature of these features~\cite{Shinar1984TheTungsten}, where Luscher~\textit{et al.}\ associate features below 18~eV to inter- or intra-band transitions~\cite{Luscher1977EnergySpectroscopy}, whereas Weaver~\textit{et al.}\ associate peaks in the region of 10 and 15~eV to a combination of overlapping surface and bulk plasmons.~\cite{Weaver1975OpticalTungsten} More specifically, using optical measurements they identify two pairs of bulk and surface plasmons at energies of 10.0 and 9.7~eV (first pair) and 15.2 and 14.8~eV (second pair), respectively. Another possible explanation as to why these plasmons are found at lower energies is that the main plasmons are damped by interband transitions.~\cite{Avery_1981} Alternatively, the lower energy bulk plasmon may only involve one group of electrons, with the main charge density from the \textit{d}-like electrons omitted in this excitation.~\cite{Weaver1974OpticalEV} Given the close proximity of the overlapping lower energy plasmons, they are often not resolved and appear as a single peak. These low energy plasmons are termed subsidiary plasmons or ``lowered plasmons'' and given that the observations by Weaver~\textit{et al.}\ appear well supported by others~\cite{Ballu1976SURFACE100, Avery_1981, Shinar1984TheTungsten}, we assign features \textbf{a} and \textbf{b} to these ``lowered'' plasmons.\par
 
The energy losses corresponding to features \textbf{d}, \textbf{e} and \textbf{f} closely match the core ionisation energies of W~4\textit{f}\textsubscript{7/2}, W~4\textit{f}\textsubscript{5/2}, and W~5\textit{p}\textsubscript{3/2}, respectively. A feature attributed to the W~5\textit{p}\textsubscript{1/2} core level should appear at approximately 40~eV, but is difficult to observe in the spectrum as it overlaps with the more intense lower energy tail of feature \textbf{g}.\par

The high energy features \textbf{g} and \textbf{h} are reported at similar energy loss positions in previous studies.~\cite{Harrower1956AugerW, Powell1958EffectsTungsten, Tharp1967EnergyTungsten,Scheibner1967InelasticSurfaces, Edwards1971EnergyTungsten, Th1979Bulk-andMetal, Luscher1977EnergySpectroscopy} Explanations regarding the origin of these features are only reported in studies by Tharp and Scheibner.~\cite{Tharp1967EnergyTungsten, Scheibner1967InelasticSurfaces} They report peaks at energies of 43~eV and 53.5~eV for tungsten similar to our work, and conclude that as no combination of surface and bulk plasmon energy loss values (i.e.\ second order plasmon) could account for these loss features, they must be attributed to interband transitions between valence band orbitals and shallow core levels. In this work, the bulk plasmon energy loss value is 23.7~eV and therefore, if two plasmons combined to form a second order plasmon, a feature should occur at 47.4~eV (23.7$\times$2 = 47.4~eV), which is approximately 5~eV higher than feature \textbf{h}. Therefore, based on the data presented here a second order plasmon is unlikely and instead these features will be termed interband transitions in the following.\par


\subsection{Core Level Photoelectron Spectroscopy and Theory}\label{XPS}

The core level spectra of tungsten offer detailed insights into the electronic structure through the presence of extended satellite features. To disentangle and identify the complex satellite structure, this work combines a large number of deep and shallow core level spectra collected with both SXPS and HAXPES and calculated from theory. Survey spectra collected with SXPS and HAXPES are presented in the Supplementary Information~VII. Additionally, the absolute binding energy (BE) of core level peaks, along with the full width at half maximum (FWHM), spin orbit splitting (SOS) separation, and relative BE separation between the main photoemission peak and corresponding satellite peaks, are listed in the Supplementary Information VIII. From the core level analysis, all core levels display asymmetric line shapes characteristic of metallic systems.~\cite{Doniach1969Many-electronMetals, Hufner_1975, Wertheim1976Many-bodyStates}\par


\subsubsection{Shallow Core Levels}\label{Shallow_CL}

The two tungsten core levels most frequently accessed with XPS are W~4\textit{f} and 4\textit{d}, as they can be easily measured using standard Al and Mg~K$\alpha$ laboratory X-ray sources. As mentioned earlier, the analysis of these core levels presents many challenges, which posses difficulties when chemical state and/or quantitative information is required. Here, high-resolution SXPS and HAXPES core level reference spectra of the shallow core levels are discussed. The information obtained from REELS is used to identify the origin and location of satellite features.\par

\begin{figure*}
\centering
    \includegraphics[keepaspectratio, width = 12.9cm]{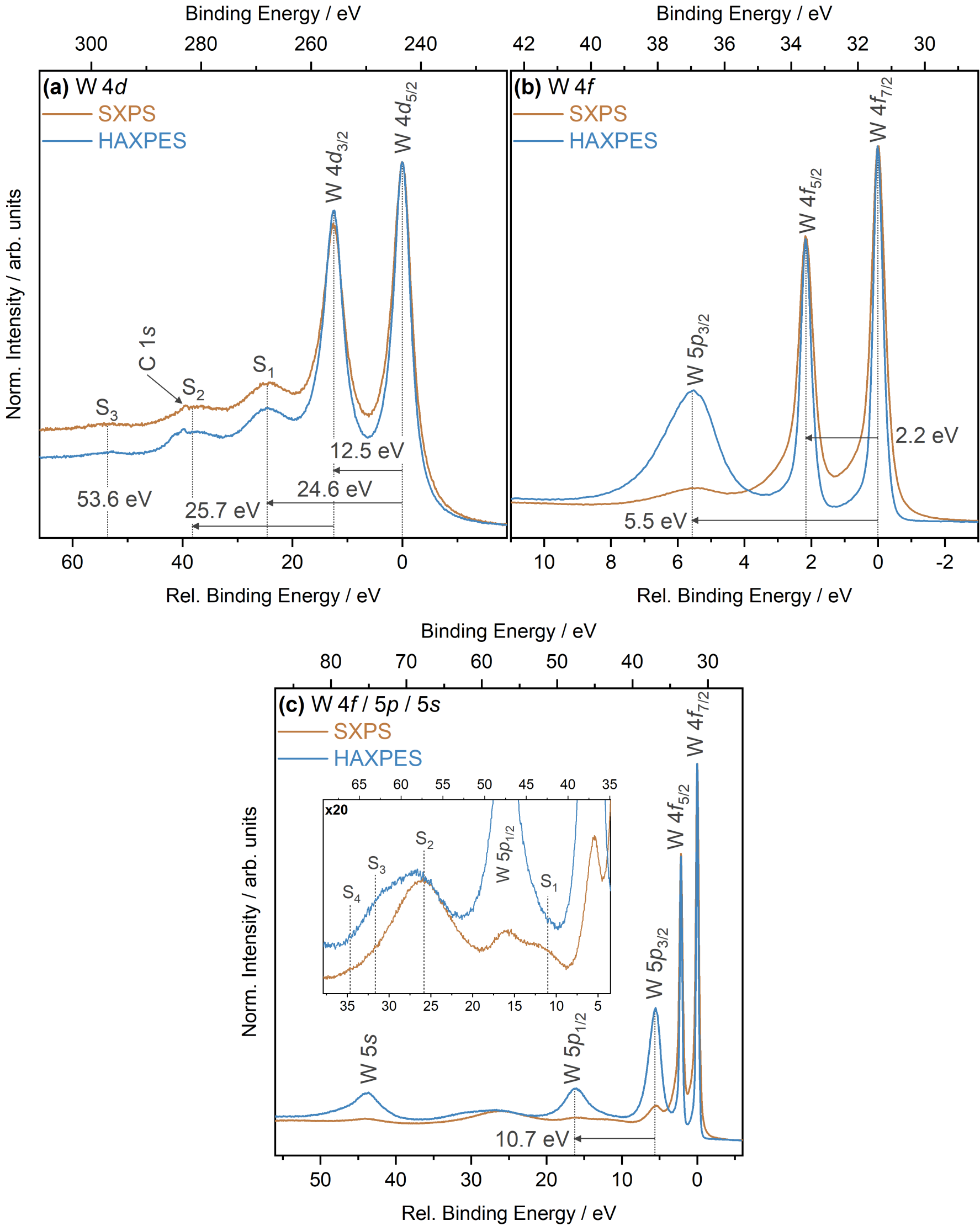}
    \caption{Shallow core level spectra collected with SXPS and HAXPES. (a) W~4\textit{d}, (b) W~5\textit{p}\textsubscript{3/2}/4\textit{f}, and (c) W~5\textit{s}/5\textit{p}/4\textit{f} core levels. The inset in (c) shows a magnified view of the satellite region. Spectra are plotted on a relative BE scale, aligning the main photoionisation peak to 0~eV. The experimental binding energy scale is also displayed above each core level spectrum.}
    \label{fig:Semi_CL}
\end{figure*}

The BE positions of the W~4\textit{f} and 4\textit{d} core level peaks are in good agreement with past measurements.~\cite{Biloen1973X-RayOxide, McGuire1973Study1, Ng1976StudiesSpectroscopy,Colton1976ElectronicSpectroscopy, Colton1978ElectrochromismSpectroscopy, Nyholm1980Core72-83, Mueller1988AOtextsubscript2,   Takano1989NitrogenationImplantation,  Engelhard2000ThirdSpectroscopy, Powell2012RecommendedSolids} The W~4\textit{d} core level (Fig.~\ref{fig:Semi_CL}(a)) displays three satellite features, labelled S\textsubscript{1}-S\textsubscript{3}, which appear in identical positions in both SXPS and HAXPES spectra. Features S\textsubscript{1} and S\textsubscript{2} are located at 24.6~eV and 25.7~eV relative to the 4\textit{d}\textsubscript{5/2} and 4\textit{d}\textsubscript{3/2} photoemission peaks, respectively, and are bulk plasmon satellites. The broad and low intensity feature S\textsubscript{3} is detected at 53.6~eV relative to the 4\textit{d}\textsubscript{5/2} photoionisation peak and based on the REELS assignments, originates from an interband transition. There is no significant difference in plasmon intensity or structure when comparing the SXPS and HAXPES spectra. This suggests that both techniques are more sensitive to the bulk plasmon, with the surface plasmon only weakly contributing to the spectra.\par 

The rapid decay of photoionisation cross sections $\sigma$ at higher excitation energies ($\sigma$~$\propto$~E\textsuperscript{-3}) is often considered an intrinsic limitation of HAXPES measurements. However, in the case of the close lying W~5\textit{s}/5\textit{p}/4\textit{f} core level the differences in the decay rates of cross sections between the orbitals can be used to aid interpretation of the spectra. A plot showing the photoionisation cross sections as a function of photon energy can be found in the Supplementary Information~VI. In the HAXPES experiment the intensity of the shallow 5\textit{p} and 5\textit{s} core level peaks is enhanced compared to that of 4\textit{f}. The 5\textit{p}\textsubscript{3/2}/4\textit{f}\textsubscript{7/2} Scofield cross section ratio for tungsten is 0.15 at a photon energy of 1.4867~keV (Al~K$\alpha$), and rises to 1.68 at 5.9267~keV.~\cite{Scofield1973TheoreticalKeV, Kalha20} The enhancement in signal intensity is clear in the experimental data in Fig.~\ref{fig:Semi_CL}(c). This enables the accurate determination of the SOS of the W~5\textit{p} doublet peaks and was found to be 10.6~eV matching closely with the value reported by Sundberg~\textit{et al.} who also used HAXPES to determine the 5\textit{p} SOS of tungsten.~\cite{Sundberg2014UnderstandingHAXPES}\par

The satellite features in the shallow W~4\textit{f}/5\textit{p}/5\textit{s} core levels are not as pronounced as those in the 4\textit{d} core level. The inset in Fig.~\ref{fig:Semi_CL}(c) highlights the satellite region between the 5\textit{p}\textsubscript{3/2} and 5\textit{s} core lines. Three satellite features appear in this region, as well as a higher BE low intensity satellite feature. Feature S\textsubscript{1} appears on the lower BE side of the 5\textit{p}\textsubscript{1/2} peak at approximately 11.2~eV relative to the main 4\textit{f}\textsubscript{7/2} core line. This feature leads to a slight asymmetric broadening of the 5\textit{p}\textsubscript{1/2} core line in the HAXPES spectrum. However, the satellite is much clearer in the SXPS spectrum, owing to the reduced 5\textit{p}\textsubscript{1/2} photoionisation cross section. To the best of our knowledge, this satellite has not been reported before and matches closely to the ``lowered'' plasmon loss energy peaks listed in Table~\ref{REELS_peak}. Two additional features, labelled S\textsubscript{2} and S\textsubscript{3} are observed at approximately 25.9~eV and 31.7~eV, respectively, relative to the 4\textit{f}\textsubscript{7/2} core line. Feature S\textsubscript{2} can be assigned to the bulk plasmon and is linked to the plasmon generation by 4\textit{f} electrons. Whilst feature S\textsubscript{2} appears in both the SXPS and HAXPES spectra, feature S\textsubscript{3} is only visually prominent in the HAXPES spectrum. The position of this feature relative to the 5\textit{p}\textsubscript{3/2} is 25.2~eV, meaning that it is the bulk plasmon loss stemming from the 5\textit{p}\textsubscript{3/2} electron. The large enhancement of its intensity in the HAXPES spectrum suggests that, much like core levels, the intensity of plasmon satellites has a dependence on the photoionisation cross sections. This further reinforces the benefit of using SXPS and HAXPES in parallel as the strategic tuning of photon energy allows for previously unidentified features to be enhanced. Lastly, feature S\textsubscript{4} is located at approximately 34.7~eV from the main photoionisation peak and has almost negligible intensity, making it difficult to observe, although it appears more prominent in the SXPS spectrum than the HAXPES. Its energy position suggests that it may be attributed to an interband transition stemming from the 4\textit{f} and/or 5\textit{p} core level electrons as the BE position occurs at an energy similar to the energy loss region of peaks \textbf{d}-\textbf{e} in the REELS spectrum.\par

\begin{figure*}[ht!]
\centering
    \includegraphics[keepaspectratio, width=17.2 cm]{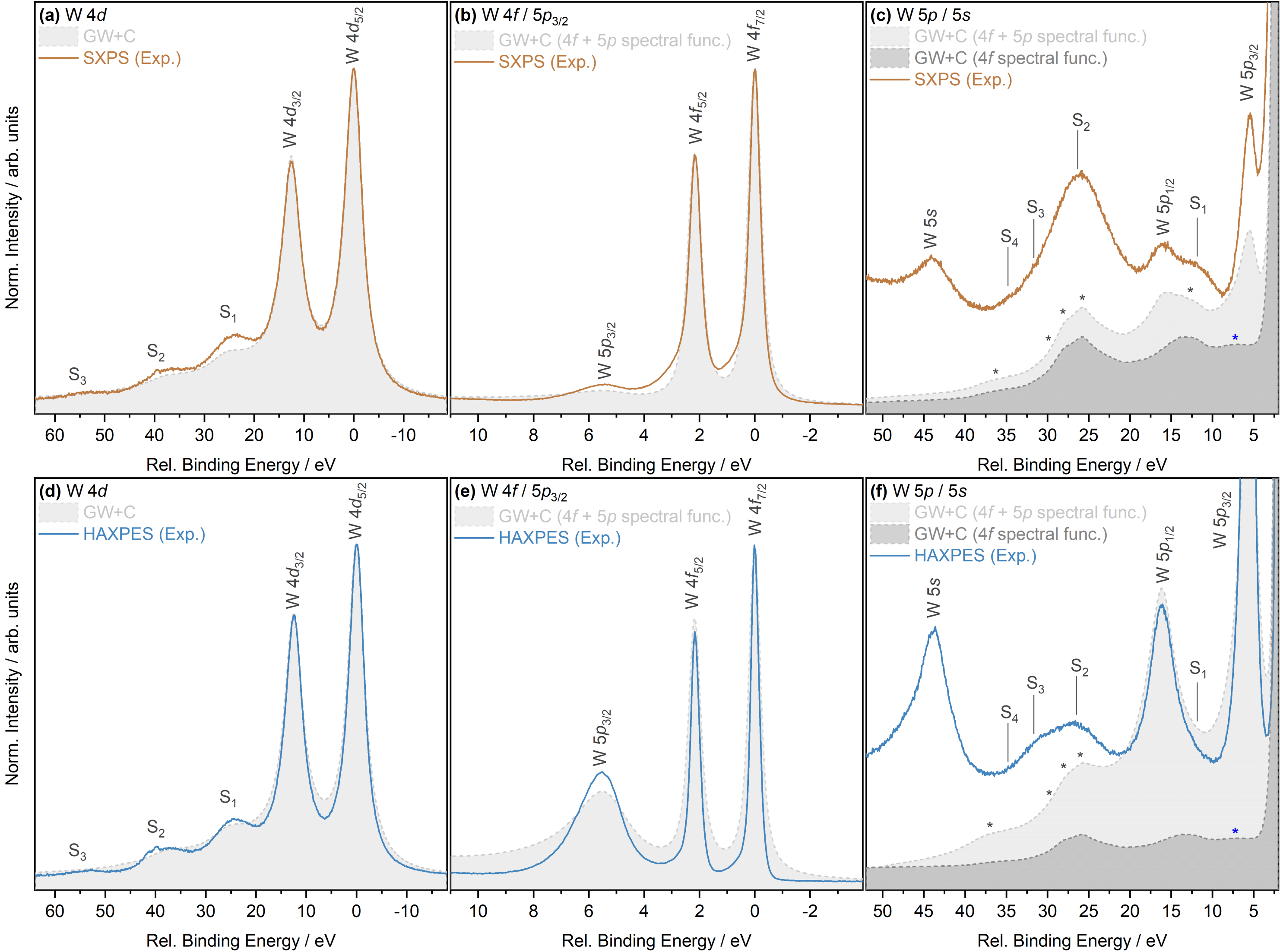}
    \caption{GW+C simulated spectra and experimental SXPS and HAXPES spectra of the shallow core levels. Going from left to right the W 4\textit{d}, W 4\textit{f} and W~5\textit{p}\textsubscript{3/2}, and enlarged W~5\textit{p}/5\textit{s} region core level spectra. The top row compares the GW+C simulated spectra to SXPS spectra, whereas the bottom row compares to HAXPES spectra. A Shirley-type background was removed from the experimental data to aid comparison to theory for the W~4\textit{d} and W~4\textit{f}/5\textit{p}\textsubscript{3/2} spectra only. To aid with identifying the satellite structure in the 5\textit{p}-5\textit{s} region shown in (c) and (d), two simulated spectra are shown. One was calculated by using a combination of the 4\textit{f} and 5\textit{p} spectral function, with the other only using the 4\textit{f} spectral function. Asterisks mark the satellite positions determined using GW+C. The black asterisks are shared between both the 4\textit{f}+5\textit{p} and 4\textit{f} GW+C simulated spectra, whereas the blue asterisk is used to label the satellite feature only visible in the 4\textit{f} GW+C simulated spectrum.}
    \label{fig:GW+shallow}
\end{figure*}

In order to gain further insights into the satellite structures observed, the W~4\textit{d} and W~4\textit{f}/5\textit{p} SXPS and HAXPES spectra are compared to GW+C simulated spectra (see Fig.~\ref{fig:GW+shallow}). As for the 4\textit{d} spectra, shown in Figs.~\ref{fig:GW+shallow}(a) and (d), good agreement between experiment and theory is observed. The GW+C approach does remarkably well in predicting the relative intensity and binding energy positions of satellites S\textsubscript{1} and S\textsubscript{2}. However, the third satellite at 53.6~eV is not captured by the calculation, likely owing to its relatively small intensity. Figs.~\ref{fig:GW+shallow}(b) and (e) display the 4\textit{f} and 5\textit{p}\textsubscript{3/2} simulated core level spectra, where again good agreement between the relative intensities and broadening of the 4\textit{f} doublet is found, especially for the SXPS case. In both the SXPS and HAXPES simulated spectra, the 5\textit{p}\textsubscript{3/2} relative intensity is not as well described, which is due to the theoretical line widths being overestimated,~\cite{Perkins1991Tables1-100} leading to a reduction in the peak height.\par

Figs.~\ref{fig:GW+shallow}(c) and (f) show an enlarged view of the 5\textit{p} and 5\textit{s} region. Four satellite features can be observed in the SXPS spectra, and the simulated spectra determined from combining the 4\textit{f} and 5\textit{p} spectral functions match well, with the individual predicted satellite features marked with asterisks. Moving from lower to higher BE values, the GW+C approach predicts five satellites, appearing at 12.7~eV, 25.8~eV, 28.0~eV, 29.7~eV and 36.5~eV. The first predicted feature at 12.7~eV correlates well with S\textsubscript{1} and relates to the ``lowered'' plasmons found in the REELS spectrum. This feature is more apparent in Fig.~\ref{fig:GW+shallow}(c) due to the reduced intensity of the 5\textit{p} core levels with SXPS. Predicted features at 25.8 and 28.0~eV overlap and their separation ($\approx$2.2~eV) matches the SOS of the 4\textit{f} core level. These features are attributed to bulk plasmons generated by the excitation of 4\textit{f}\textsubscript{7/2} and 4\textit{f}\textsubscript{5/2} core electrons, and contribute to the satellite S\textsubscript{2} seen in the experimental spectra. Overall, the GW+C predicted features describe the experimental observed satellite feature S\textsubscript{2} in both SXPS and HAXPES spectra well. S\textsubscript{3} at 30.9~eV, which is only clearly visible in the HAXPES spectrum due to cross section enhancement, and is attributed to a plasmon associated with the excitation of the 5\textit{p}\textsubscript{3/2} electrons, is difficult to observe in the simulated spectra due to the limitations of the theoretical line widths used~\cite{Perkins1991Tables1-100} and the resulting smearing of features. The last remaining predicted feature at 36.5~eV is very low in intensity, but can be assigned to the weak satellite S\textsubscript{4} visible in the SXPS spectrum. This satellite most likely arises from interband transitions between the 4\textit{f} and conduction band states. It is expected to have higher intensity in the SXPS spectrum due to the enhanced photoionisation cross section compared to HAXPES.\par

In order to disentangle contributions from the 4\textit{f} and 5\textit{p} spectral functions Figs.~\ref{fig:GW+shallow}(c) and (f) also display the simulated spectra determined using only the 4\textit{f} spectral functions. In the 4\textit{f} only simulation, all satellites previously described are present and an additional low intensity, low BE feature satellite is visible at approximately 7.3~eV from the 4\textit{f}\textsubscript{7/2} peak. Whereas in the 4\textit{f} and 5\textit{p} simulation but also the experimental spectrum, this feature is difficult to observe as it sits underneath the 5\textit{p}\textsubscript{3/2} core line. This 7.3~eV predicted feature does not appear in the REELS spectrum reported here, but Weaver~\textit{et al.}\ suggest that features within this region relate to interband transitions between valence and conduction band states.~\cite{Weaver1975OpticalTungsten} The reason why this feature is not observed in our REELS spectrum is likely due to the employment of a high electron incident energy, which creates a large inelastic background, masking the low intensity feature.\par

\subsubsection{Evaluation of Core Level Line Widths}

From the discussion presented so far it is clear that the intrinsic complexity of the shallow core lines, both in metallic tungsten and exacerbated when oxide and other compound states are present, complicates analysis. Therefore, a clear motivation exists to explore other core levels where these constraints are not present and hard X-rays can be used to unlock additional higher energy core levels.\par

An important aspect when combining core level spectra of different orbital natures and binding energies is the difference in lifetime broadening. The ideal alternative core level to analyse would be well separated from other neighbouring core levels, and has a natural line width similar to that of the W~4\textit{f} core level, as this is narrow enough to resolve chemical shifts. It is important to remember, that core level line widths in XPS have both a Lorentzian and Gaussian component, with the former attributed to lifetime $\tau$ broadening effects in response to the creation of a core hole during the photoemission process and the latter attributed to non-lifetime effects (e.g.\ instrumental factors, temperature, phonon broadening etc.). The Lorentzian contribution to the line shape is given by,
\begin{equation}\label{equ:Lorentz}
\centering
    I(E) = I(E_0)\frac{\Gamma^2}{(E-E_0)^2 + \Gamma^2}\textrm{,}
\end{equation}
where $I$ is the spectral intensity at a given energy $E$, $E_0$ is the centroid energy of the Lorentzian peak, and $2\Gamma$ is the natural line width (e.g.\ core hole lifetime broadening), with $\Gamma$ given by
\begin{equation}\label{equ:Life}
\centering
    \Gamma = \hbar/\tau = (6.582\times10^{-16}~eV{\cdot}s)/\tau \textrm{.}
\end{equation}

\begin{figure}
\centering
    \includegraphics[keepaspectratio, width=\linewidth]{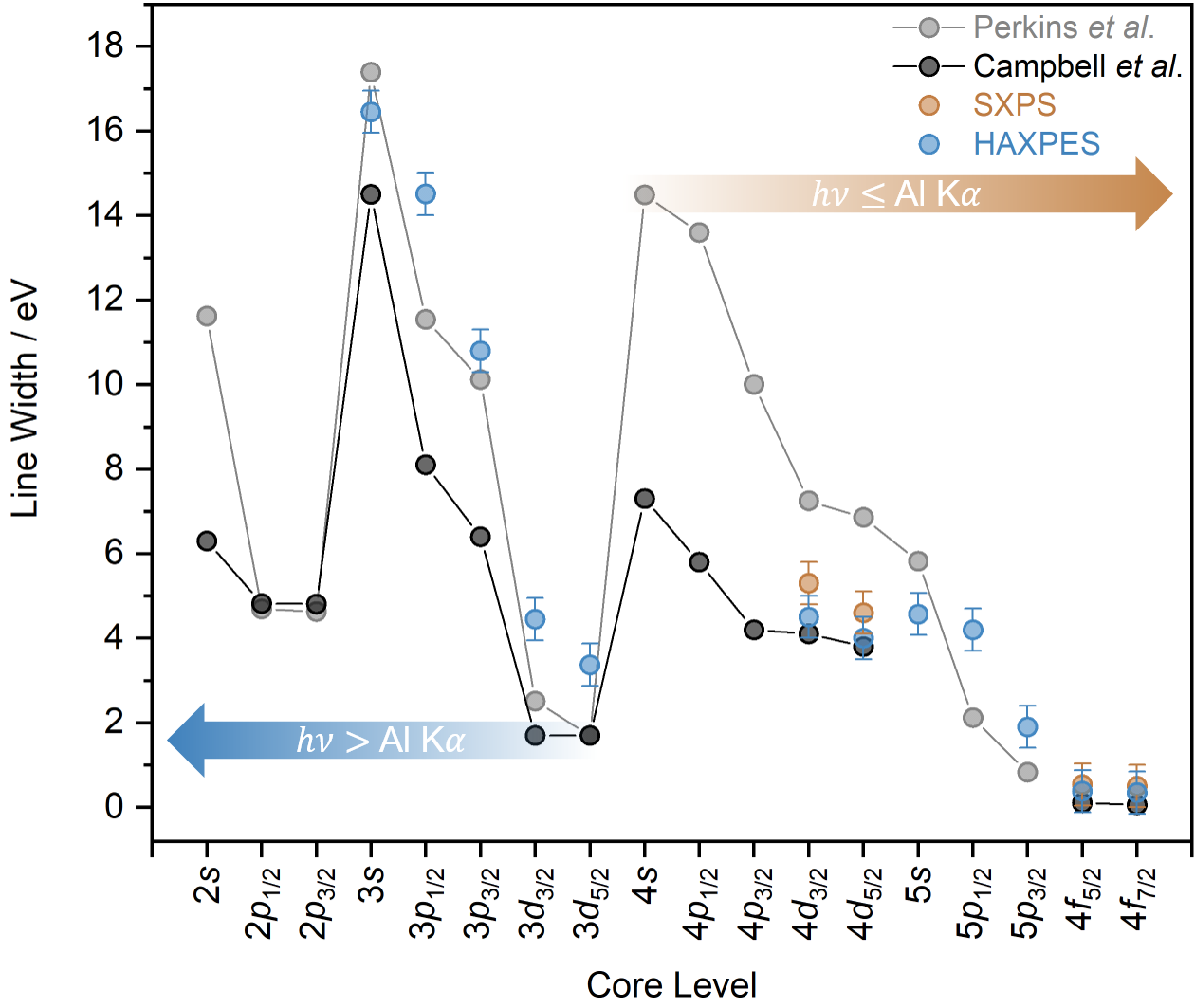}
    \caption{Comparison of the total core line widths of SXPS and HAXPES data with reported natural line-width values from Refs.~\cite{Perkins1991Tables1-100,Campbell2001WidthsLevels}. The W~5\textit{p} and W~5\textit{s} core level line widths are not reported by Campbell~\textit{et~al.}\ and Perkins~\textit{et~al.}\ do not calculate the W~4\textit{f} line widths. All core levels up to and including the W~4\textit{s} can be accessed with the Al~K$\alpha$ photon energy, whereas to access deeper core levels (3\textit{d} and above), a photon energy higher than Al~K$\alpha$ is required. }
    \label{fig:FWHMs}
\end{figure}

Fig.~\ref{fig:FWHMs} compares the measured FWHMs of all core levels using both SXPS and HAXPES to the theoretical natural line widths reported by Perkins~\textit{et al.},~\cite{Perkins1991Tables1-100} which include the sum of both radiative and non-radiative line widths, along with line widths determined from the comparison of available experimental and theoretical data by Campbell~\textit{et al.}, which they define as ``recommended line widths''.~\cite{Campbell2001WidthsLevels} The core levels vary substantially in their line widths with the 3\textit{s} core level having the largest (16.5~eV) and the 4\textit{f} having the smallest (0.4~eV), both determined from the HAXPES measurements. The main trend observed is that with increasing angular momentum (i.e.\ going from \textit{s} to \textit{d} orbitals) the natural line width decreases (e.g.\ $\Gamma$(3\textit{s})$>$ $\Gamma$(3\textit{p}\textsubscript{1/2}) $>$ $\Gamma$(3\textit{p}\textsubscript{3/2}) $>$ $\Gamma$(3\textit{d}\textsubscript{3/2}) $>$ $\Gamma$(3\textit{d}\textsubscript{5/2})). This can be attributed to a reduction in the Coster-Kronig-Auger decay.~\cite{Fuggle_1980} The measured shallow (4\textit{f}-4\textit{d}) core level line widths are in better agreement with the recommended line widths, whereas the deeper (3\textit{s}-3\textit{d}) core levels show a better agreement with the theoretical values. Campbell~\textit{et al.}\ note a scarcity of available data for the deeper core levels of elements above Z~=~55, with only X-ray emission spectroscopy (XES) data available rather than XPS. Additionally, the SXPS recorded line widths for the 4\textit{f} and 4\textit{d} core lines are broader than the HAXPES recorded line widths due to the better energy resolution of HAXPES. Due to the low intensity of the 5\textit{p} and 5\textit{s} core lines in the SXPS spectra (as will be shown in Section~\ref{Shallow_CL}) the accurate determination of their FWHM was not possible. Fuggle~\textit{et al.}\ suggest that the differences are due to other non lifetime-broadening effects.~\cite{Fuggle_1980} Whereas Ohno~\textit{et al.}\ attribute certain discrepancies due to the theory approach taken by Perkins~\textit{et al.}, suggesting the many-body-theory approach is necessary to offer better line width prediction to the experimental results.~\cite{Ohno, Fuggle_1980}\par

From the HAXPES measurements, a FWHM of 3.4~eV for the W~3\textit{d}\textsubscript{5/2} core line and a large 3\textit{d} SOS of 62.1~eV are found, both of which are sufficient to allow for chemical shifts to be resolved. Additionally, Fig.~\ref{fig:FWHMs} shows that its natural line width and therefore Lorentzian contribution is lower than that of the 4\textit{d} core lines, and therefore is advantageous from an analytical perspective. Based on this information, the 3\textit{d} core level can be considered as an alternative to the shallow 4\textit{d} and 4\textit{f} core levels and can be used to provide additional complimentary information from a different depth perspective.\par
\subsubsection{Deep Core Levels}

The need to access deeper core levels for tungsten has not seen as much interest compared to titanium and silicon, where the Ti~1\textit{s} and Si~1\textit{s} core level is frequently accessed with HAXPES in favour of the Ti~2\textit{p} and Si~2\textit{p} core levels, as analysis is more straightforward due to the lack of impeding satellite structures, the higher photoionisation cross sections, and the absence of SOS effects to consider.~\cite{Panaccione2012HardProperties, Church2015UnderstandingSpectroscopy, Regoutz2018ASystem} However, in light of the observation that the 3\textit{d} core level may offer complimentary information to the commonly used shallow core levels, there is clear motivation to explore deeper core lines. Given the current popularity of HAXPES, there are also greater opportunities to conduct such experiments.~\cite{Kalha2021Hard2020} Therefore, the following discussion reports the first in-depth description of the deep core levels of tungsten metal collected with HAXPES.\par

\begin{figure*}[t]
\centering
    \includegraphics[keepaspectratio, width =12.9 cm]{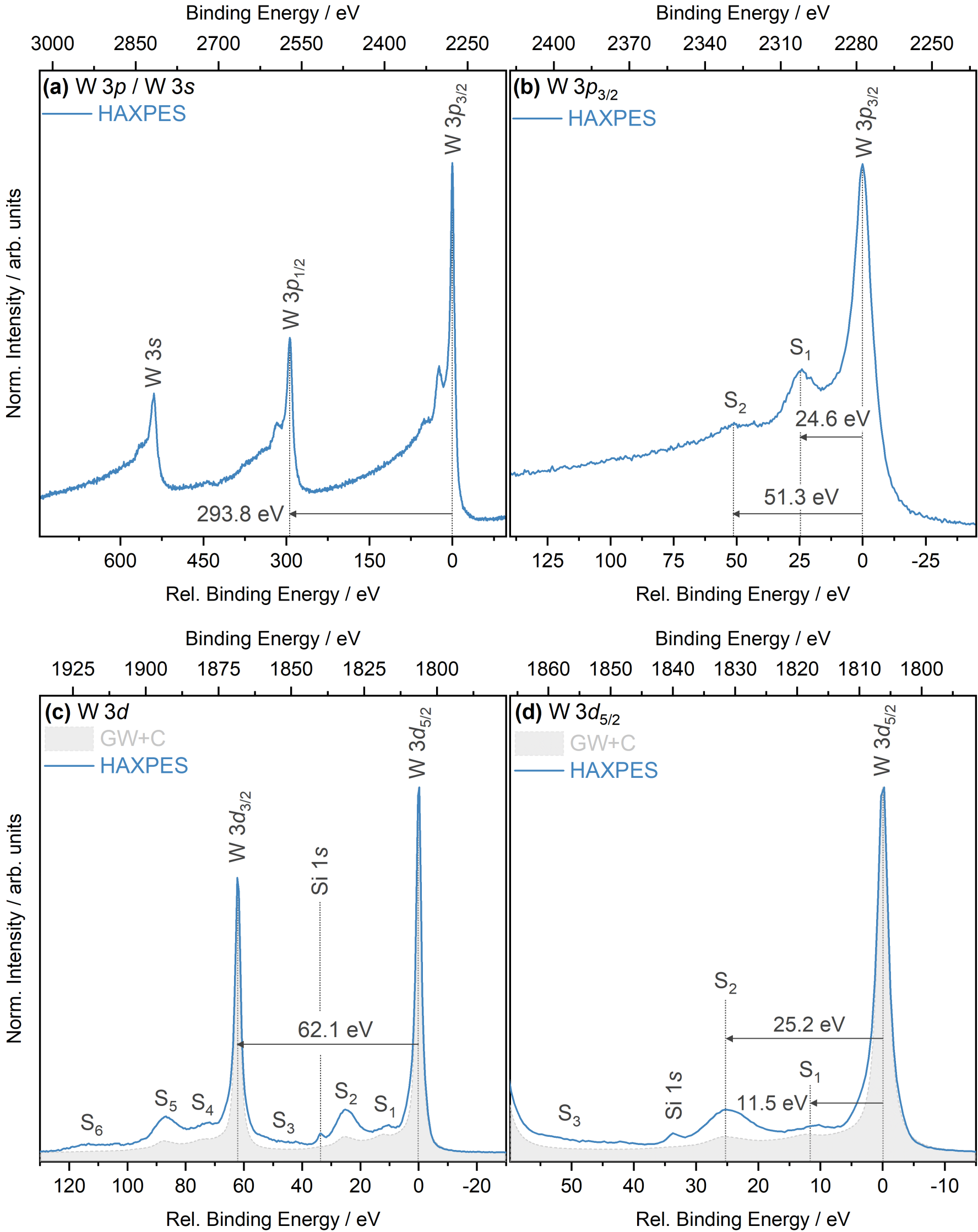}
    \caption{Core level spectra from HAXPES experiments and theoretical GW+C results for W~3\textit{d}. (a) W~3\textit{p} and W~3\textit{s}, (b) W~3\textit{p}\textsubscript{3/2}, (c) W~3\textit{d}, and (d) W~3\textit{d}\textsubscript{3/2} core levels. Spectra are plotted on a relative BE scale, aligning the main photoionisation peak to 0~eV. The experimental binding energy scale is also displayed above each core level spectrum. The Si~1\textit{s} core line appearing in the W~3\textit{d} core level originated from the ex-situ sample preparation. (c) and (d) also show the comparison between the HAXPES W~3\textit{d} core level spectrum and the GW+C simulated spectrum. A Shirley-type background was removed from the experimental spectrum and both the simulated and experimental spectra were normalised to the maximum peak intensity to aid with the comparison.}
    \label{fig:HX_Cl}
\end{figure*}

Fig.~\ref{fig:HX_Cl} displays the W~3\textit{p}/3\textit{s}, W~3\textit{p}\textsubscript{3/2}, W~3\textit{d} and W~3\textit{d}\textsubscript{5/2} core levels, which are only accessible using excitation energies above those of conventional soft X-ray laboratory sources. They display a complex satellite structure with large decaying backgrounds on the higher BE side of the main photoionisation peaks. To the best of our knowledge, there have only been two reported studies on the W~3\textit{d} core level but none on the W~3\textit{s} or W~3\textit{p} core levels of tungsten metal. Wagner~\cite{Wagner1978X-rayEnergy} appears to be the first to access the 3\textit{d}\textsubscript{5/2} core line using a Au~M$\alpha$ (2.123~keV) photon source, reporting a BE position of 1807.6~eV. However, no spectra were displayed in this study. More recently, Sundberg~\textit{et al.}\ accessed the 3\textit{d} core level using HAXPES ($h\nu$ = 3 and 6~keV), reporting a BE position of the 3\textit{d} doublet core lines of 1806.8~eV and 1868.9~eV.~\cite{Sundberg2014UnderstandingHAXPES} Sundberg~\textit{et al.}\ only show the 3\textit{d}\textsubscript{5/2} core line, and satellite features were not captured.\par

The BE positions of the 3\textit{d}\textsubscript{5/2} and 3\textit{d}\textsubscript{3/2} core level peaks observed in the current HAXPES experiment are 1806.3~eV and 1868.4~eV, respectively, with a SOS of 62.1~eV, matching closely with the values reported by Sundberg~\textit{et al.} Additionally, the BE position of the 3\textit{s}, 3\textit{p}\textsubscript{1/2} and 3\textit{p}\textsubscript{3/2} core lines are 2817.3~eV, 2571.3~eV, and 2277.5~eV, respectively, with the 3\textit{p} doublet having a SOS of 293.8~eV. The 3\textit{p}\textsubscript{3/2} core level, displayed in Fig.~\ref{fig:HX_Cl}(b), displays two satellite features located at 24.6~eV (S\textsubscript{1}) and 51.3~eV (S\textsubscript{2}) relative to the main photoionisation peak, and these features are mirrored by the 3\textit{p}\textsubscript{1/2} spin component and also the 3\textit{s} core level. To identify the origin of these satellites, comparison to the REELS data is helpful. The satellite features occur at similar positions to the energy loss values of features \textbf{c} and \textbf{h} (see Fig.~\ref{fig:REELS}). Following the assignments made for the REELS data, S\textsubscript{1} corresponds to the bulk plasmon and S\textsubscript{2} to an interband transition. The small differences in the BE positions of the features compared to the reported energy loss values in the REELS measurement can be attributed to the differences in the underlying excitation mechanisms between the two techniques. Plasmon satellites in photoemission experiments differ from plasmon-related features in electron energy loss experiments as they contain both intrinsic (due to photo-excitation) and extrinsic (electron scattering during transport to surface) plasmon losses, whereas energy loss experiments only contain the latter.~\cite{Bates_1979, Inglesfield1981PLASMONPHOTOEMISSION} Consequently, this can lead to a difference in intensity of these features, which can influence the accurate determination of their positions. Additionally, lifetime broadening effects in SXPS/HAXPES can impede the accurate determination of weak satellite features that lie close to the main photoemission peak.\par

The W~3\textit{d} core level displayed in Fig.~\ref{fig:HX_Cl}(c) displays six satellite features (S\textsubscript{1}-S\textsubscript{6}), shared equally and mirrored by each spin component (S\textsubscript{1} = S\textsubscript{4}, S\textsubscript{2} = S\textsubscript{5}, S\textsubscript{3} = S\textsubscript{6}). Satellite features S\textsubscript{1} and S\textsubscript{2} in the 3\textit{d}\textsubscript{5/2} core level region, shown in Fig.~\ref{fig:HX_Cl}(d), appear at relative BE positions of 11.5~eV and 25.2~eV, respectively. S\textsubscript{2} and S\textsubscript{5} are the most intense satellite features and by using the previous assignments in REELS are assigned to the bulk plasmon. Features S\textsubscript{1} and S\textsubscript{4} appear at a similar position to the energy loss position of peak \textbf{a} in the REELS spectrum and therefore are attributed to a subsidiary plasmon. Whilst S\textsubscript{6} in the 3\textit{d}\textsubscript{3/2} region is clearly observed, the mirrored feature S\textsubscript{3} is hard to distinguish due to the impeding lower BE tail of the 3\textit{d}\textsubscript{3/2} peak. S\textsubscript{6} occurs at ca. 52.2~eV relative to the 3\textit{d}\textsubscript{3/2} peak, matching closely to the second satellite feature in the 3\textit{p}\textsubscript{3/2} core level, and are therefore considered to be an interband transition.\par

As for the shallow core levels, theoretical GW+C results were used to gain a better understanding of the complex satellite features observed. Fig.~\ref{fig:HX_Cl}(c) and (d) displays the simulated W~3\textit{d} core level, calculated using GW+C, and provides a direct comparison to the HAXPES data. Good agreement is observed between theory and experiment, with the core level line widths, line shape, and relative intensities being well reproduced. This suggests that both the applied Scofield photoionisation cross sections~\cite{Scofield1973TheoreticalKeV} and recommended line width values determined by Campbell~\textit{et al.}~\cite{Campbell2001WidthsLevels} work well for the case of tungsten.

Moreover, the use of the 4\textit{f} spectral function to simulate this deep 3\textit{d} core level is effective and shows that this approach could be used to simulate deep core levels for other metallic elements. In terms of the prediction of the satellite peaks, the GW+C approach is able to describe the first two satellite features. The satellites are located at 11.5~eV and 25.2~eV in the experiment, which agrees well with the theory positions of 12.5~eV and 25.3~eV. The second satellite is under-predicted in intensity in the simulated GW+C spectrum relative to the experimental spectrum. As mentioned above, plasmons are classified into either intrinsic or extrinsic categories, with both contributing to the experimental spectrum. However, GW+C only describes the intrinsic losses, which is why a reduction in signal intensity is observed. The third satellite feature (S\textsubscript{3}, S\textsubscript{6}) does not appear clearly in the GW+C calculation, similar to the case of the 4\textit{d} simulated spectra (Figs.~\ref{fig:GW+shallow}(a) and (d)) and again is most likely due to its vanishingly small intensity relative to the background.

\subsubsection{Comparison of Core Level Satellites}

\begin{figure}[ht!]
\centering
    \includegraphics[keepaspectratio, width=\linewidth]{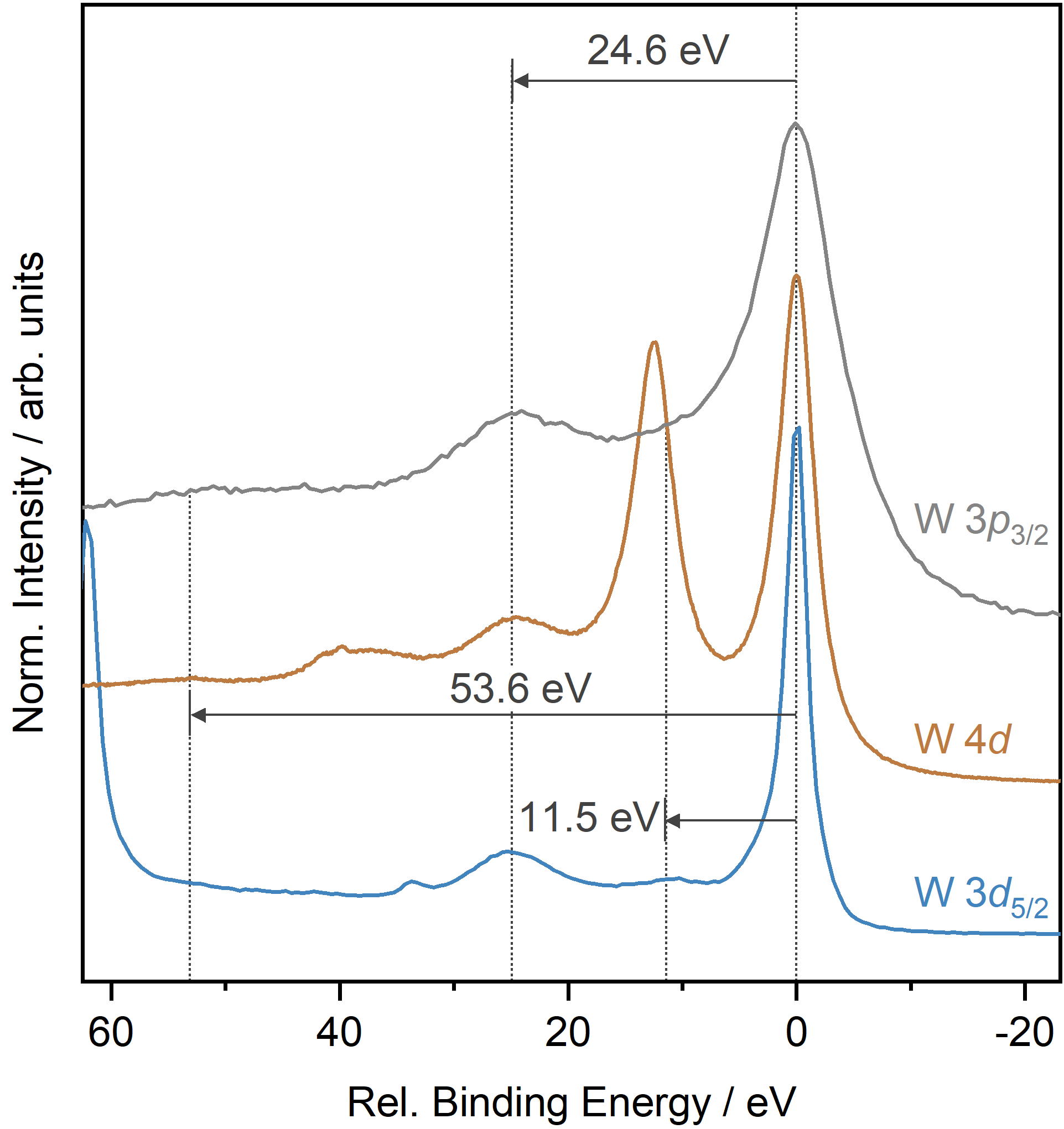}
    \caption{W~3\textit{p}\textsubscript{3/2}, W~4\textit{d}, and W~3\textit{d}\textsubscript{5/2} core level HAXPES spectra. Spectra are offset vertically, normalised to their maximum intensity, and aligned relative to the main photoionisation peak.}
    \label{fig:Satellite_Comparison}
\end{figure}

When comparing all core level spectra, similarities in their satellite features become clear. This is expected, as based on the above discussion, these features are a fingerprint of the intrinsic electronic structure of tungsten. Fig.~\ref{fig:Satellite_Comparison} shows the comparison between the W~3\textit{p}\textsubscript{3/2}, W~3\textit{d}\textsubscript{5/2}, and W~4\textit{d} core level HAXPES spectra. It is strikingly clear that all core level spectra (including the shallow W~4\textit{f}/5\textit{p} core level) share the same bulk plasmon satellite feature located at ca. 25~eV from the main photoionisation peak. Moreover, 3\textit{p}\textsubscript{3/2} and 4\textit{d} share the same low intensity satellite at ~53.6~eV. The 3\textit{d}\textsubscript{5/2} spectrum on the other hand displays a low intensity satellite at 11.5~eV, which is also present on the lower BE side of the 5\textit{p}\textsubscript{1/2} core line (See Fig.~\ref{fig:Semi_CL}(c)). Due to the SOS of the 4\textit{d} core level, this satellite feature will appear under the 4\textit{d}\textsubscript{3/2} core line, which is why it has never been observed. Additionally, it will also fall under the higher BE tail of the 3\textit{p}\textsubscript{3/2} core line. However, given the low intensity of the satellite observed in the 3\textit{d} core level, the presence of the satellite in the 4\textit{d} core level will most likely not need to be considered during peak-fit analysis of the region. Likewise, given the presence of the 53.6~eV satellite in the 4\textit{d} and 3\textit{p}\textsubscript{3/2} core level regions, one can assume it is also present in the 3\textit{d}\textsubscript{5/2} region, but due to its low intensity and close proximity to the 3\textit{d}\textsubscript{3/2} core line it is smeared out.\par

Uncovering hidden satellite feature such as the ones discussed here, highlights the benefit of using both HAXPES and SXPS to understand the detailed satellite structures in core level spectra. A similar approach was used by Woicik~\textit{et al.}, who discovered the appearance of a low intensity 5~eV satellite hidden underneath the Ti~2\textit{p}\textsubscript{1/2} core line of SrTiO\textsubscript{3} by comparing the spectrum to the deeper Ti~1\textit{s} core level.~\cite{Woicik_SrTiO3}

\subsection{Valence Electronic Structure}\label{sec:Electronic}

\begin{figure}[ht]
\centering
    \includegraphics[keepaspectratio, width=\linewidth]{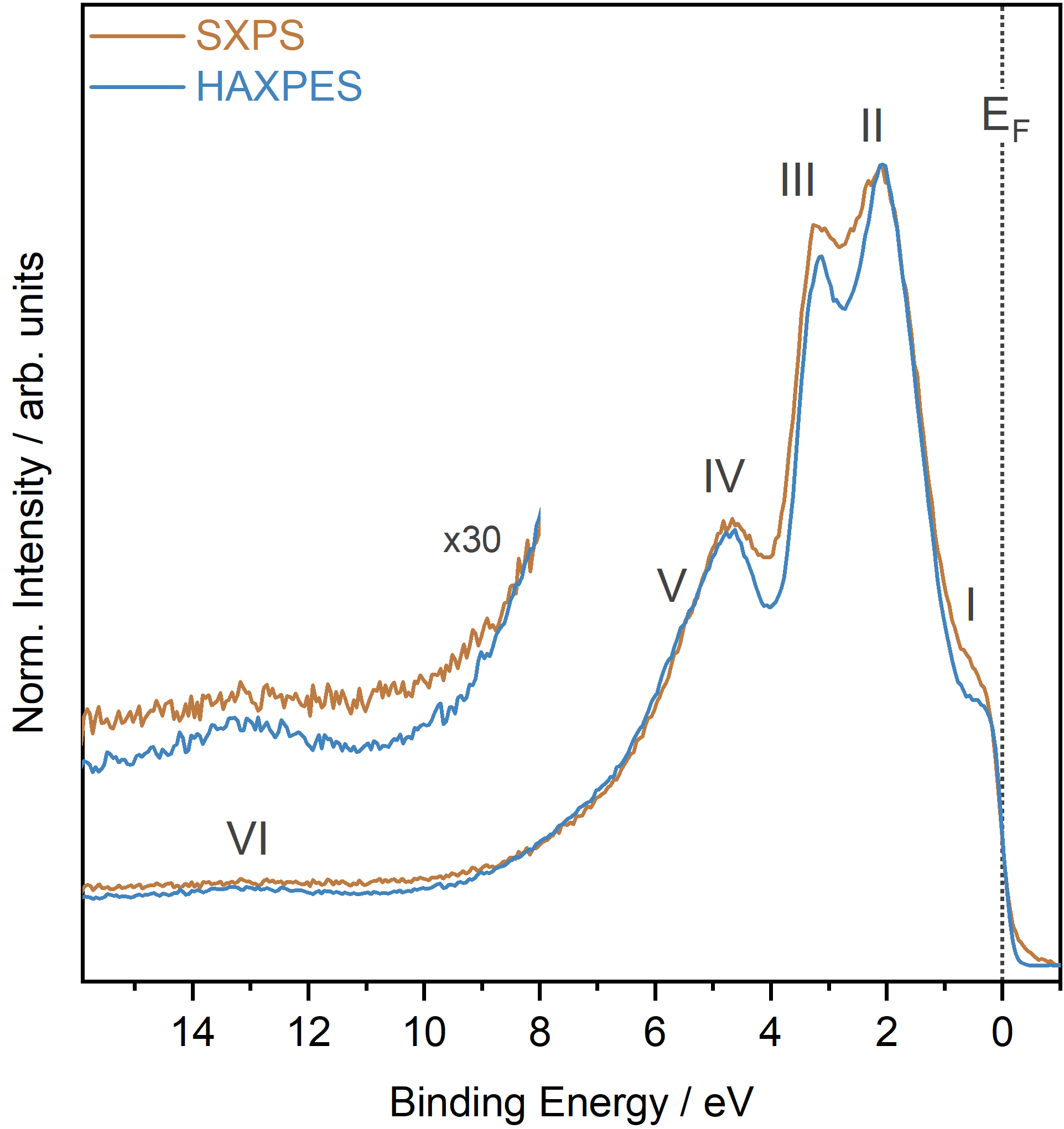}
    \caption{High resolution SXPS and HAXPES valence band spectra. Spectra are normalised to the maximum intensity after the removal of a constant linear background. The inset shows a $\times$30 magnification of feature VI plotted on the same x-axis scale.}
    \label{fig:VB_exp}
\end{figure}

The information on the electronic structure gained from REELS and core level PES can be further extended by considering the valence electronic structure of tungsten. Several studies report the valence band spectrum of tungsten metal, predominantly focusing on PES using soft or ultra-violet photon energies, constraining the measurements to the sample surface.~\cite{Penchina1974PhotoemissionFilms, Colton1976ElectronicSpectroscopy, Engelhard2000ThirdSpectroscopy} Additionally, to date, no study has used HAXPES to capture a bulk valence band spectrum of tungsten that can be directly compared to theory. To address these limitations, high resolution valence band spectra with improved signal-to-noise ratio were obtained using both SXPS and HAXPES. Fig.~\ref{fig:VB_exp} displays the collected valence band spectra and shows that the same features are present with both SXPS and HAXPES, and appear in near identical BE positions, which is expected for a metallic system. The subtle differences observed between the SXPS and HAXPES spectra are due to a combination of different energy resolution as well as differences in photoionisation cross sections.\par

Six key features are identified and labelled with Roman numerals -- I, II, III, IV, V and VI -- located at approximately 0.4~eV, 2.0~eV, 3.2~eV, 4.7~eV, 5.5~eV and 13.0~eV, respectively. The general shape of the valence band SXP spectrum is in good agreement with previous studies.~\cite{Penchina1974PhotoemissionFilms, Colton1976ElectronicSpectroscopy, Fleisch1982AnWO3, Engelhard2000ThirdSpectroscopy} This work is able to present a much higher resolution spectrum, resolving features II and III, where previous studies fail.~\cite{Penchina1974PhotoemissionFilms, Engelhard2000ThirdSpectroscopy} The BE positions of features I-IV match closely to those presented by Hussain~\textit{et al.}\ who reported similar features at 0.6~ eV, 2.3~eV, 3.2~eV and 4.8~eV using angle-resolved PES ($h\nu$ = Al~K$\alpha$).~\cite{Hussain1980Temperature-dependentEffects} Feature V is more apparent in the HAXPES spectrum due to the subtle difference in cross sections between the 5\textit{d} and 6\textit{p} states. 

Feature VI has not been observed to date. It appears close to features \textbf{a} and \textbf{b} reported in the REELS spectrum, which are attributed to the ``lowered'' plasmon losses. Feature VI is visible in both SXPS and HAXPES spectra, excluding a pure surface phenomenon. These observations give weight to the argument that this feature is an intrinsic part of the electronic structure of tungsten. A similar feature is also observed above the valence band of other BCC transition metals,~\cite{Werfel1983PhotoemissionOxides, Buabthong2017VanadiumXPS} which exhibit such ``lowered'' plasmons, however, this feature is never discussed.~\cite{Weaver1973OpticalEV, Weaver1974OpticalEV} A similar observation was shared by {{\L}}awniczak-Jab{{\l}}o{\'n}ska~\textit{et al.}\ who highlighted the presence of a low intensity valence band feature at 12~eV for molybdenum, much like feature VI in our spectra for tungsten.~\cite{Lawn_1988} They attribute the feature to an energy loss associated with an interband transition, which again further reinforces the assumption made earlier that feature VI is intrinsic to the electronic structure of tungsten and is due to what Weaver~\textit{et al.}\ states is a ``lowered'' plasmon loss.\par

\begin{figure*}[ht]
\centering
    \includegraphics[keepaspectratio, width=17.2 cm]{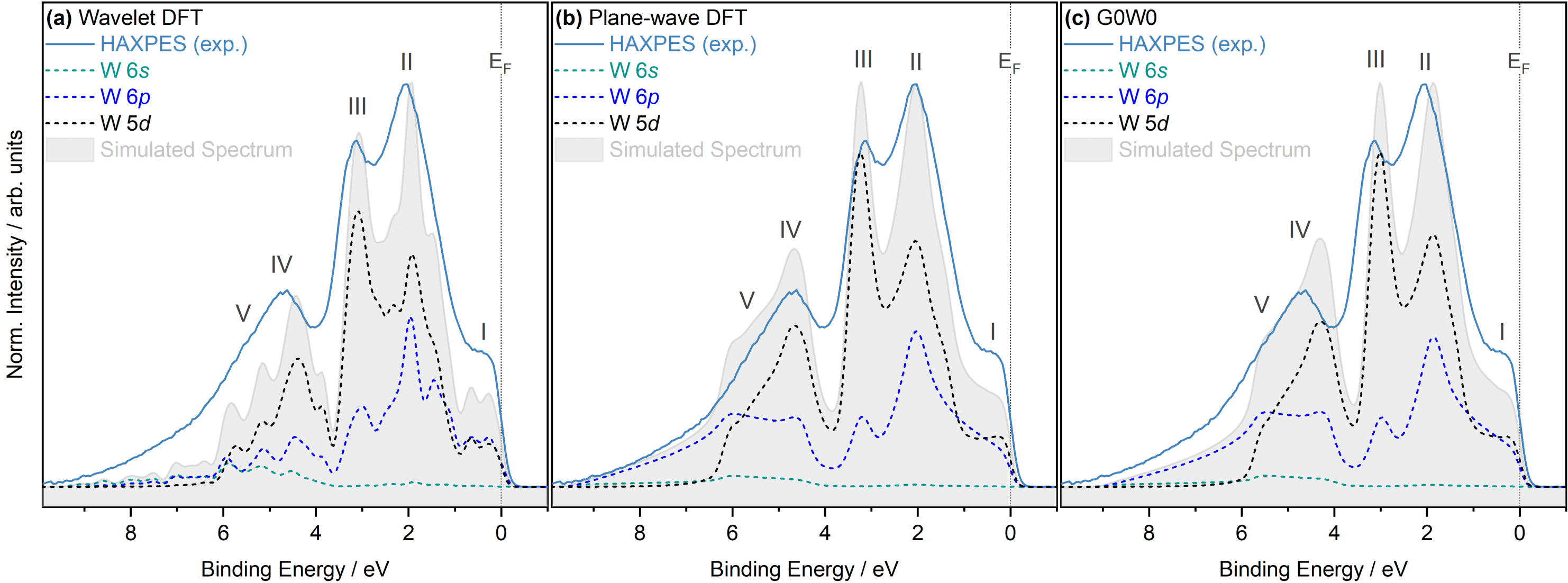}
    \caption{Comparison of simulated PDOS spectra calculated using DFT and G0W0 approaches with the HAXPES valence band spectra, including the comparison with (a) DFT using a wavelet basis set, (b) DFT using a plane-wave basis set, and (c) G0W0. The contributions to all PDOS spectra displayed are photoionisation cross section weighted using the ``optimised'' method outlined in Section~\ref{X_section_method}. A Shirley-type background was removed from the experimental spectrum to allow direct comparison to the theory. The PDOS contributions have also been suitably broadened to match the experimental broadening. Simulated Spectrum refers to the sum of the weighted PDOS. See Supplementary Information~IX and X for the G0W0 unweighted PDOS and comparison of PDOS to the SXPS collected valence band spectrum, respectively.}
    \label{fig:Theory_Comp}
\end{figure*}

Fig.~\ref{fig:Theory_Comp} displays the calculated PDOS from both DFT approaches (using the wavelet and plane-wave basis sets) along with the G0W0 approach, and compares them to the experimental HAXPES valence band spectrum. The PDOS shown have been weighted using the Scofield W~5\textit{d} and W~6\textit{s} values at the given photon energy as well as the W~6\textit{p} cross section determined using the ``optimised'' approach. All three theory approaches are in good agreement with the experimental result, clearly identifying all key features and their relative energy positions. The wavelet DFT approach (i.e.\ using LS-DFT) shows remarkable similarity to the plane-wave DFT approach (i.e.\ using conventional cubic-scaling DFT). However, the plane-wave DFT is smoother, which is also the case in the unweighted total densities of states (TDOS) (see Supplementary Information~XI). Additional calculations were performed to assess the influence of both the lattice parameter and the pseudopotential choice, including whether or not the 5\textit{s} and 5\textit{p} states are included in the core or treated as valence states. The PDOS was found to be insensitive to both of these parameters, while spin orbit coupling was also found to have little influence. Thus, the differences between the TDOS for the two DFT approaches can be attributed to the smaller effective $k$-point sampling of the wavelet-based results. The strong effect of the choice of $k$-point sampling (or equivalently, supercell size) can be clearly seen in the Supplementary Information~IV, where the calculated PDOS are presented for different supercell sizes. Due to the high computational cost, 3456 atoms was the largest system attempted, and although it appears close to convergence it could be interesting to consider larger supercells in future work.

Aside from the differences coming from the TDOS, the plane-wave DFT approach, shown in Fig.~\ref{fig:VB_exp}(b) projects a greater contribution from the 6\textit{p} states between 6-9~eV, whereas the projection from the wavelet DFT approach (Fig.~\ref{fig:VB_exp}(a)) shows that the contributions from all states is almost minimised in this region. Referring again to the unweighted PDOS, the two DFT approaches show some difference in the relative contributions coming from both the 6\textit{s} and 6\textit{p} states below 4~eV, while the contribution arising from the 5\textit{d} states is similar in both approaches. Upon applying the photoionisation cross sections, the stronger 6\textit{p}-contribution in the plane-wave projection leads to higher relative peak heights in this region, while in the wavelet case the larger 6\textit{s}-contribution is lost due to the smaller 6\textit{s} cross section, leading to smaller relative peak heights. In other words, the differences between the two approaches below 4~eV are solely due to difference in the TDOS, i.e.\ resulting from the different $k$-point sampling, while the differences above 4~eV arise due to \textit{both} the differences in the TDOS and in the projection scheme. Nonetheless, both DFT approaches give a good description of the experimental results, thereby highlighting the viability of using LS-DFT for the modelling of disordered metal alloy systems in large supercells. Whereas, comparing the plane-wave DFT PDOS to the G0W0 PDOS minimal differences are observed in the shape of the projection, with the only difference being that the G0W0 predicts a narrower band width. The overestimation of the band width from DFT calculations on metallic systems is a well known problem and is often a motivation for using G0W0.~\cite{Northrup1989QuasiparticleMetals, Qiu2020ComparisonMonolayer}\par

The theory reflects the expected electronic structure with spatially localised 5\textit{d} states providing the majority contribution to the valence band and free-electron like 6\textit{sp} bands only giving a small contribution. Feature I is shown to arise from a mixing of the 6\textit{p} and 5\textit{d} states with both showing an equal contribution. Similarly, features II and III also arise from a mixing of 5\textit{d} and 6\textit{p} states, however, the 5\textit{d} states dominate, especially for feature III. Feature IV also arises predominantly from mixing of the 5\textit{d} states with 6\textit{p} states with a small contribution from 6\textit{s} states. This is also the case for feature V, which appears as a distinct shoulder on feature IV, which has a marginally higher contribution from \textit{s} states.\par

\begin{figure}[ht!]
\centering
    \includegraphics[keepaspectratio, width=\linewidth]{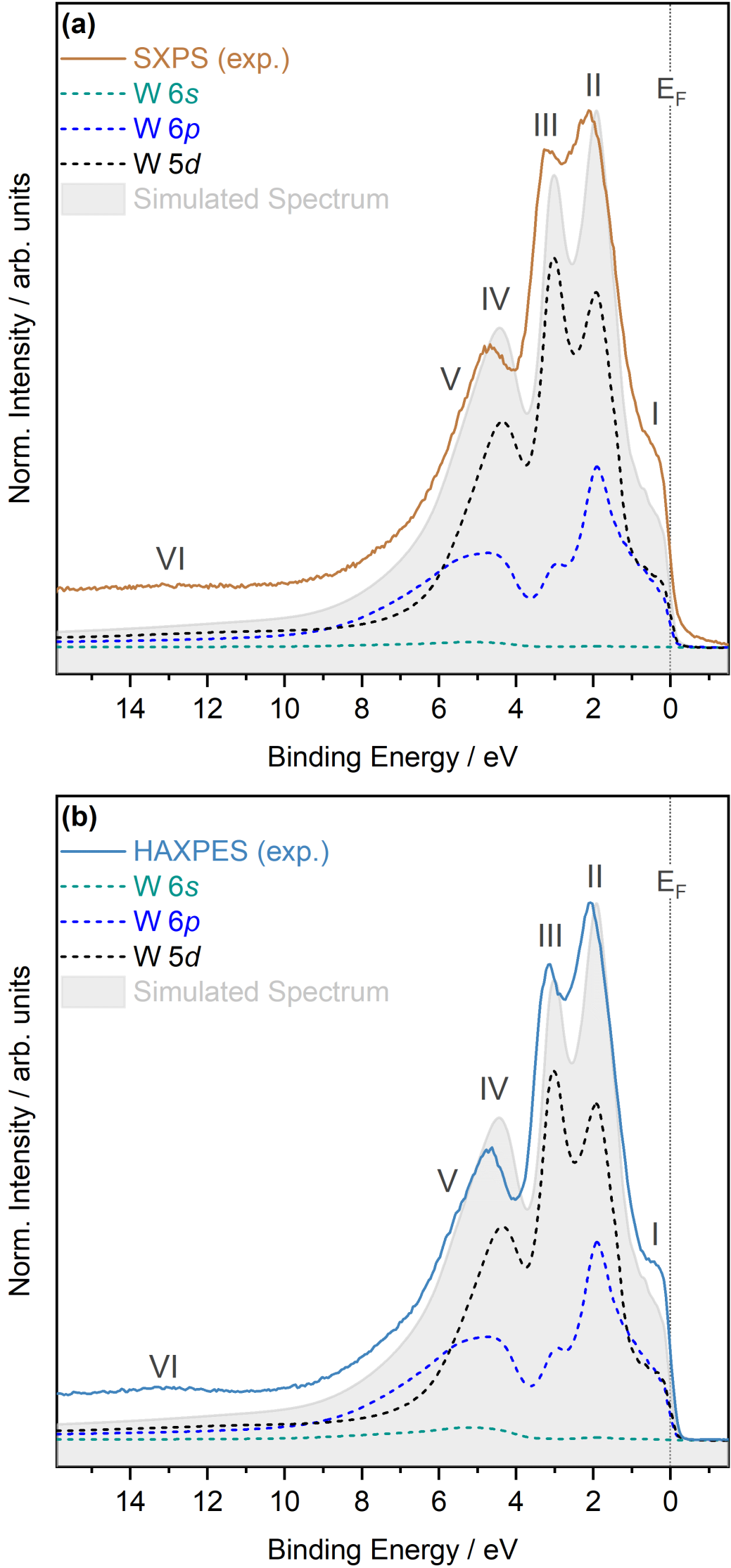}
    \caption{Comparison of the simulated PDOS spectrum calculated using the GW+C approach to the valence band collected with (a) SXPS and (b) HAXPES. The PDOS contributions have been cross section weighted according to the ``optimised'' method outlined in Section~\ref{X_section_method}, and as the GW+C approach considers lifetime broadening, no additional broadening was applied. Simulated Spectrum refers to the sum of the weighted PDOS. See the Supplementary Information~IX for the GW+C unweighted PDOS.}
    \label{fig:GW_C_VB}
\end{figure}

All three theory approaches match the BE positions of the higher BE region (IV and V) well, but struggle to accurately describe the overall shape of the valence band in this region. Such discrepancies between DFT and photoelectron spectroscopy (PES) have been noted by others,~\cite{Strocov1998AbsoluteCu, Marini2002QuasiparticleApproximation, Panaccione_2005} who have put forward two possible reasons -- firstly the necessity of approximating exact exchange and correlation potentials, and secondly, the fundamental difference between DFT and PES, in that DFT is only considering the ground state, whereas PES reflects additional final state effects. For these reasons, the inclusion of self-energy corrections (i.e.\ G0W0 and GW+C) are typically needed to generate an improved comparison of theory to experiment. Therefore, the GW+C approach was used, with the result displayed in Fig.~\ref{fig:GW_C_VB}, which clearly provides a better agreement with the experimental spectra, especially in capturing the shape of the region around features IV and V. The GW+C PDOS curves that have been constructed from the calculated spectral functions contain the effects of lifetime broadening as well as photoemission satellites. In contrast, the DFT and G0W0 PDOS curves do not contain these effects. The ability of the GW+C method to more accurately predict the shapes of features IV and V in the experimental spectrum indicates that lifetime broadening has a significant influence on the appearance of these peaks.\par

\section{Conclusions}

This work presents a comprehensive analysis of the relationship between the electronic structure and features in the photoelectron spectrum of tungsten metal, combining state-of-the-art experimental and theoretical approaches. 

First, exploration of REELS data enabled the identification of a large number of spectral features related to plasmons and interband transitions. Insights gathered from REELS were used to inform the identification of satellite features in PES core state spectra. In-depth analysis of both SXPS and HAXPES core level spectra provides new insights into the nature of specific transitions underlying the observed satellite features, with spectral functions calculated from GW+C underpinning the experimental assignments. Direct comparisons between the shallow and deep core levels allowed for the identification of hidden, previously not identified satellite features. Cross section effects and the opportunity to access alternative core levels to the commonly used but complex 4\textit{f} and 4\textit{d} core levels clearly demonstrate the impact of HAXPES experiments. The deep 3\textit{d} core level offers a relatively narrow FWHM and small Lorentzian contribution, making it a feasible alternative to the shallow core levels. The core level data are further completed by a detailed investigation of the valence band of tungsten using experiment as well as employing multiple levels of theory. LS-DFT was successfully applied and showed good agreement to conventional cubic-scaling DFT, enabling future studies on large, complex, disordered, multi-metallic systems modelled with LS-DFT with high accuracy and reproducibility.\par
The present results offer critical insights for both fundamental studies and for crucial scientific and industrial applications involving tungsten, laying the foundation for the exploration of its nanostructures and compounds, as well as device-relevant heterostructures. Finally, the strategy presented here allows future exploration of other transition metals, which have similarly complex photoelectron spectra and electronic structure.

\begin{acknowledgments}
CK acknowledges the support from the Department of Chemistry, UCL. NKF acknowledges support from the Engineering and Physical Sciences Research Council (EP/L015277/1). AR acknowledges the support from the Analytical Chemistry Trust Fund for her CAMS-UK Fellowship.
LER acknowledges support from an EPSRC Early Career Research Fellowship (EP/P033253/1). JL and JMK acknowledge funding from EPSRC under Grant No. EP/R002010/1 and from a Royal Society University Research Fellowship (URF/R/191004). This work used the ARCHER UK National Supercomputing Service via JL’s membership of the HEC Materials Chemistry Consortium of UK, which is funded by EPSRC (EP/L000202). JJGM and SM acknowledge the support from the FusionCAT project (001-P-001722) co-financed by the European Union Regional Development Fund within the framework of the ERDF Operational Program of Catalonia 2014-2020 with a grant of 50\% of total cost eligible, the access to computational resources at MareNostrum and the technical support provided by BSC (RES-QS-2020-3-0026). Part of this work was carried out using supercomputer resources provided under the EU-JA Broader Approach collaboration  in the Computational Simulation Centre of International Fusion Energy Research Centre (IFERC-CSC). 
\end{acknowledgments}

\section*{Disclosures}
The authors declare no conflicts of interest.

\section*{Data Availability}
The data that support the findings of this study are available within the article and its supplementary material. Freely accessible versions of all survey, core level and valence band spectra collected with SXPS and HAXPES are available on Figshare at \url{https://doi.org/10.6084/m9.figshare.16432617}. All data concerning the BigDFT calculations are openly available in the NOMAD repository at \url{https://dx.doi.org/10.17172/NOMAD/2021.08.27-1}. Any further supporting data including the Quantum Espresso DFT and GW+C calculations are available from the corresponding author upon reasonable request.

\section*{Supplementary Information}
Further details regarding the experimental and computational methods can be found in the Supplementary Information. Additionally, full tabulated data of all values derived from the SXPS and HAXPES core level spectra can also be found in the Supplementary Information. Lastly, a detailed overview of the method and parameters used to construct the simulated core levels and determine the optimum photoionisation cross section weighting, variants of the LS-DFT PDOS calculations using different atom numbers, and survey spectra can also be found in the Supplementary Information.

\bibliography{references.bib}
\bibliographystyle{apsrev4-1}

\end{document}


\preprint{APS/123-QED}

\title[PRB]{Lifetime effects and satellites in the photoelectron spectrum of tungsten metal: Supplementary Information}

\author{C.~Kalha}
\affiliation{Department of Chemistry, University College London, 20 Gordon Street, London, WC1H~0AJ, United Kingdom.}

\author{L.~E.~Ratcliff}
\affiliation{Department of Materials, Imperial College London, London, SW7 2AZ, United Kingdom}
\affiliation{Thomas Young Centre for Theory and Simulation of Materials, London, UK}

\author{J.~J.~Gutiérrez Moreno}
\affiliation{Barcelona Supercomputing Center (BSC), C/ Jordi Girona 31, 08034 Barcelona, Spain.}

\author{S.~Mohr}
\affiliation{Barcelona Supercomputing Center (BSC), C/ Jordi Girona 31, 08034 Barcelona, Spain.}
\affiliation{Nextmol (Bytelab Solutions SL), C/ Roc Boronat 117, 08018 Barcelona, Spain.}

\author{M.~Mantsinen}
\affiliation{Barcelona Supercomputing Center (BSC), C/ Jordi Girona 31, 08034 Barcelona, Spain.}
\affiliation{ICREA, Pg. Lluís Companys 23, 08034 Barcelona, Spain.}

\author{N.~K.~Fernando}
\affiliation{Department of Chemistry, University College London, 20 Gordon Street, London, WC1H~0AJ, United Kingdom.}

\author{P.~K.~Thakur}
\author{T.-L.~Lee}
\affiliation{Diamond Light Source Ltd., Harwell Science and Innovation Campus, Didcot, Oxfordshire, OX1 3QR, United Kingdom.}

\author{H.-H.~Tseng}
\author{T.~S.~Nunney}
\affiliation{Thermo Fisher Scientific, Surface Analysis, Unit 24, The Birches Industrial Estate, East Grinstead, West Sussex, RH19 1UB, United Kingdom.}

\author{J.~M.~Kahk}
\affiliation{Department of Materials, Imperial College London, London, SW7 2AZ, United Kingdom}
\affiliation{Thomas Young Centre for Theory and Simulation of Materials, London, UK}
\affiliation{Institute of Physics, University of Tartu, W. Ostwaldi 1, 50411 Tartu, Estonia}

\author{J.~Lischner}
\affiliation{Department of Materials, Imperial College London, London, SW7 2AZ, United Kingdom}
\affiliation{Thomas Young Centre for Theory and Simulation of Materials, London, UK}

\author{A.~Regoutz}
 \email{a.regoutz@ucl.ac.uk}
\affiliation{Department of Chemistry, University College London, 20 Gordon Street, London, WC1H~0AJ, United Kingdom.}

\date{\today}
\maketitle

\newpage

 \tableofcontents

\cleardoublepage

\section{Sample Preparation} \label{prep}

The tungsten foil was cut to an approximate size of 5$\times$5~mm\textsuperscript{2} before cleaning with propan-2-ol. Removal of surface adsorbates and oxide over layers was conducted in-situ via argon ion sputtering prior to all REELS, SXPS, and HAXPES measurements until the C~1\textit{s} and O~1\textit{s} signals were minimised. Due to experimental time limitations, ex-situ polishing using silicon carbide abrasive paper was required prior to any in-situ sputtering for the HAXPES measurement only. For all measurements, the foil was secured to sample holders using adhesive, conductive carbon tape. Prior to REELS measurements, the surface was argon sputtered cleaned with a focused Ar\textsuperscript{+} ion gun operating with a 3~keV voltage for tens of minutes, rastering over a 2~$\times$~2~mm\textsuperscript{2} area. Prior to SXPS measurements, a focused 1~keV Ar\textsuperscript{+} ion gun voltage for 30~min was used, followed by a 15~min sputter cycle at 2~keV prior to measurements, rastering over a 2~$\times$~2~mm\textsuperscript{2} area. Prior to HAXPES measurements the sample was sputtered using an Ar\textsuperscript{+} ion gun operating at 2-2.5~keV accelerating voltage and 10 mA current emission. For all measurements data was collected at the centre of the sputtering area.

\begin{figure*}[ht!]
\centering
    \includegraphics[keepaspectratio, width=0.8\textwidth]{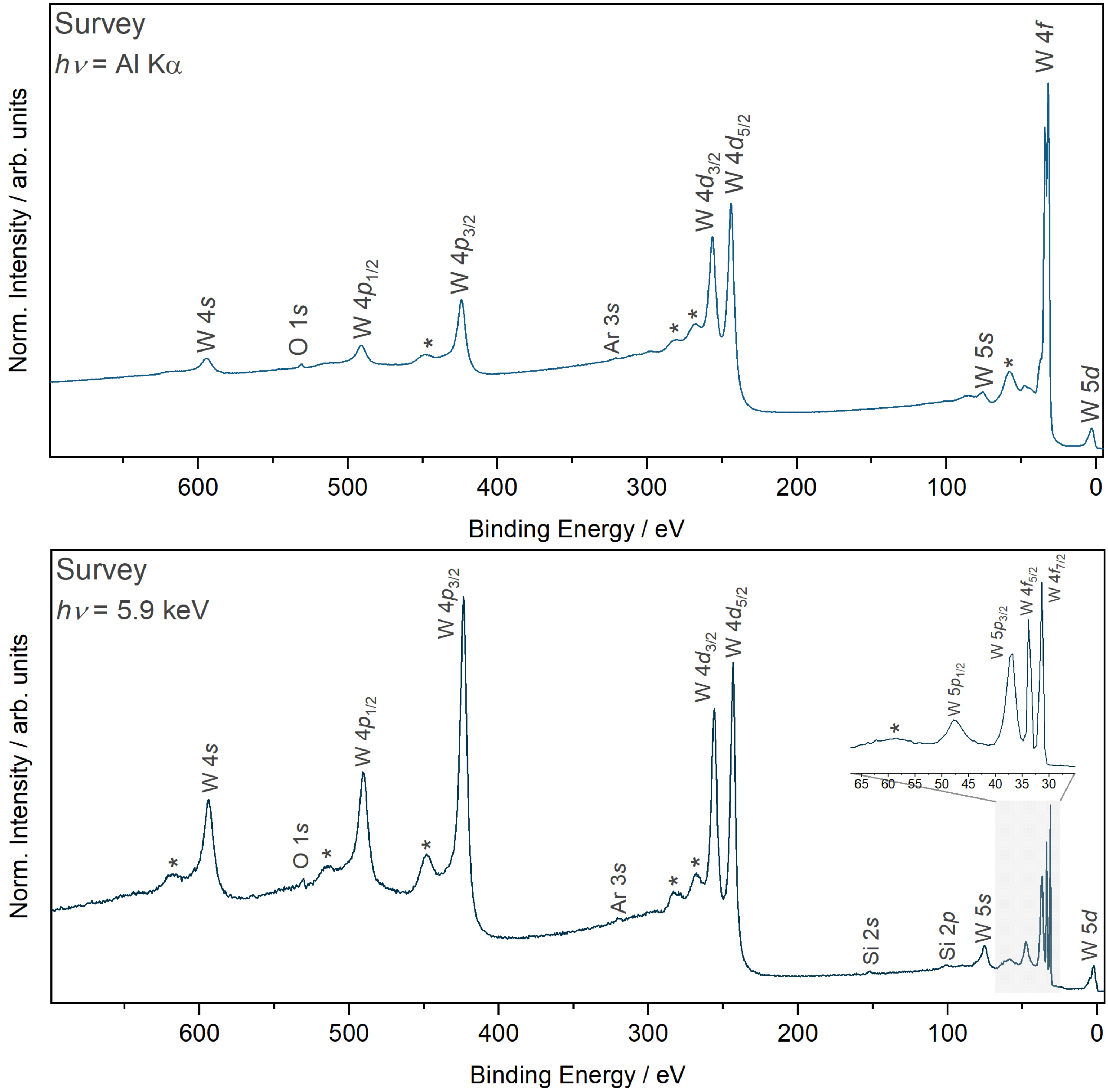}
    \caption{Survey Spectra collected with (top) SXPS and (bottom) HAXPES. All core level lines are indicated and prominent satellite features are labelled with an asterisk. The inset in the HAXPES spectrum shows a magnified view of the W 4\textit{f}/5\textit{p} region.}
    \label{fig:Survey}
\end{figure*}

The SXPS and HAXPES survey spectra are shown in Fig.~\ref{fig:Survey}. They confirm metallic tungsten with the C~1\textit{s} signal almost completely removed and the O~1\textit{s} signal minimised as much as possible. Additional silicon peaks are observed in the HAXPES spectrum at approximately 100~eV (Si~2\textit{p}), 150~eV (Si~2\textit{s}), and 1840~eV (Si~1\textit{s}), which are attributed to the polishing medium used to remove the thick native oxide over-layer. Lastly, argon implantation from the in-situ preparation method is detected, with the Ar~3\textit{s} peak at approximately 319~eV present, albeit with a very low intensity.

\cleardoublepage

\section{Energy Resolution} \label{Res}

\begin{figure*}[ht!]
\centering
    \includegraphics[keepaspectratio, width=0.8\textwidth]{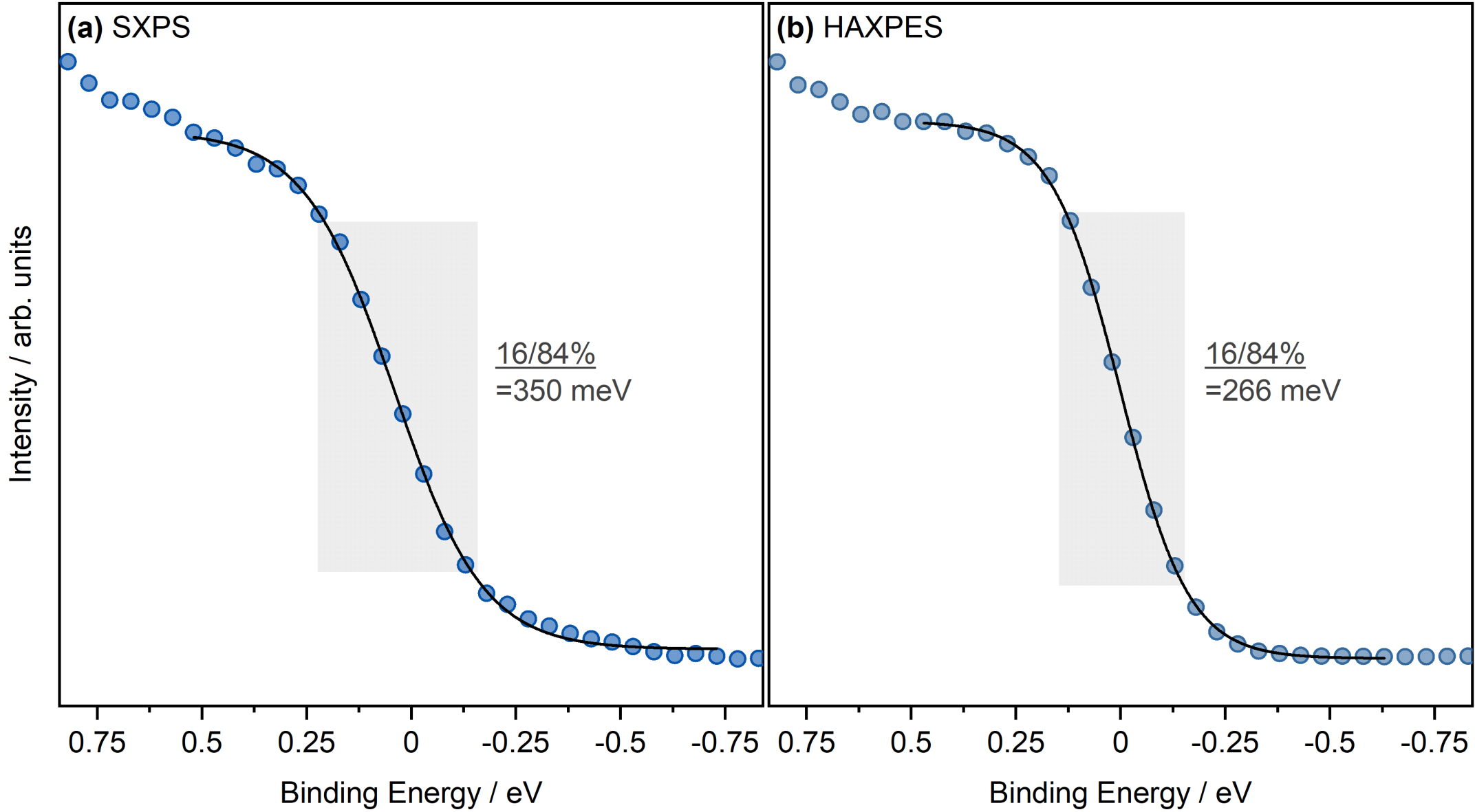}
    \caption{Fermi edge of the tungsten foil measured using (a) SXPS and (b) HAXPES including a Gaussian-broadened Fermi-Dirac distribution fit to determine the 16/84 energy resolution.}
    \label{fig:Resolution}
\end{figure*}

\cleardoublepage

\section{Additional Computational Details}\label{Computational_Details}

Quantum Espresso calculations used a scalar-relativistic norm-conserving Vanderbilt pseudopotential with 28 valence electrons per atom (W~4\textit{f}, 5\textit{s}, 5\textit{p}, 6\textit{s}, and 5\textit{d}),~\cite{Hamann2013OptimizedPseudopotentials} with an 80~Rydberg (1088~eV) wave-function cutoff. Gaussian occupation smearing of 0.01~Ry was applied at the Fermi level. The eigenstates calculated using the norm-conserving pseudopotentials were also used as inputs for GW calculations (in the main paper). In order to obtain the projections of the Kohn-Sham eigenstates onto local atomic orbitals, a separate set of DFT calculations using PAW (projector augmented wave) pseudopotentials from PSLibrary~\cite{Dal_2014} were performed, using otherwise the same parameters as above. These projections were subsequently used for constructing the DFT, GW, and GW+C projected densities of states. Calculations using two separate pseudopotentials were performed, because BerkeleyGW (in the main paper) requires the use of norm-conserving pseudopotentials, but local basis functions for obtaining projected densities of states were not available for the Vanderbilt norm-conserving pseudopotentials from Ref.~\cite{Hamann2013OptimizedPseudopotentials}. It was verified that the total densities of states obtained using the two separate pseudopotentials were very similar.\par
For the G0W0 calculations, frequencies from 0 to 60~eV were sampled on a uniform grid with a spacing of 0.08~eV, and frequencies between 60 and 400~eV were sampled with a non-uniform grid, where each successive frequency step is increased by 0.15~eV. 160 bands were included in the calculation, and a cutoff of 20~Rydberg (272~eV) was used for the dielectric matrix. Additionally, the self energy was evaluated on a frequency grid that ranges from 34.9~eV below the mean-field Fermi level to 5.1~eV above the mean-field Fermi level with a spacing of 0.25~eV. \par

BigDFT calculations employed a Krack HGH pseudopotential,~\cite{Krack2005PseudopotentialsFunctionals} which accounts for the W~6\textit{s} and 5\textit{d} states as valence electrons. A grid spacing of 0.38~bohr (0.20~{\AA}) was used. Nine support functions were employed per atom (including the unoccupied W~6\textit{p} states) with localisation radii of 7.5~bohr (3.97~{\AA}). The support functions were optimised in a 686 atom supercell then used as a fixed basis for the 3456 atom supercell, using the pseudo-fragment approach available in BigDFT~\cite{Ratcliff2019}. Calculations were performed using the Fermi Operator Expansion (FOE) approach,~\cite{Goedecker1994EfficientDynamics, Goedecker1995Tight-bindingOrbitals} implemented via the CheSS library,~\cite{Mohr2017EfficientLibrary}, since in the case of tungsten it has previously been shown to be the fastest approach for systems greater than around 1000~atoms, while also showing excellent agreement with standard cubic scaling DFT.~\cite{Mohr2018LinearBasis} The cutoff for the density kernel was set to 11.0~bohr (5.82~{\AA}). A finite electronic temperature of 0.005~{Ha} (0.136~eV) was employed.

\cleardoublepage

\section{Supercell Convergence}\label{Supercell_Convergence}

\begin{figure*}[ht!]
\centering
    \includegraphics[keepaspectratio, width=0.75\textwidth]{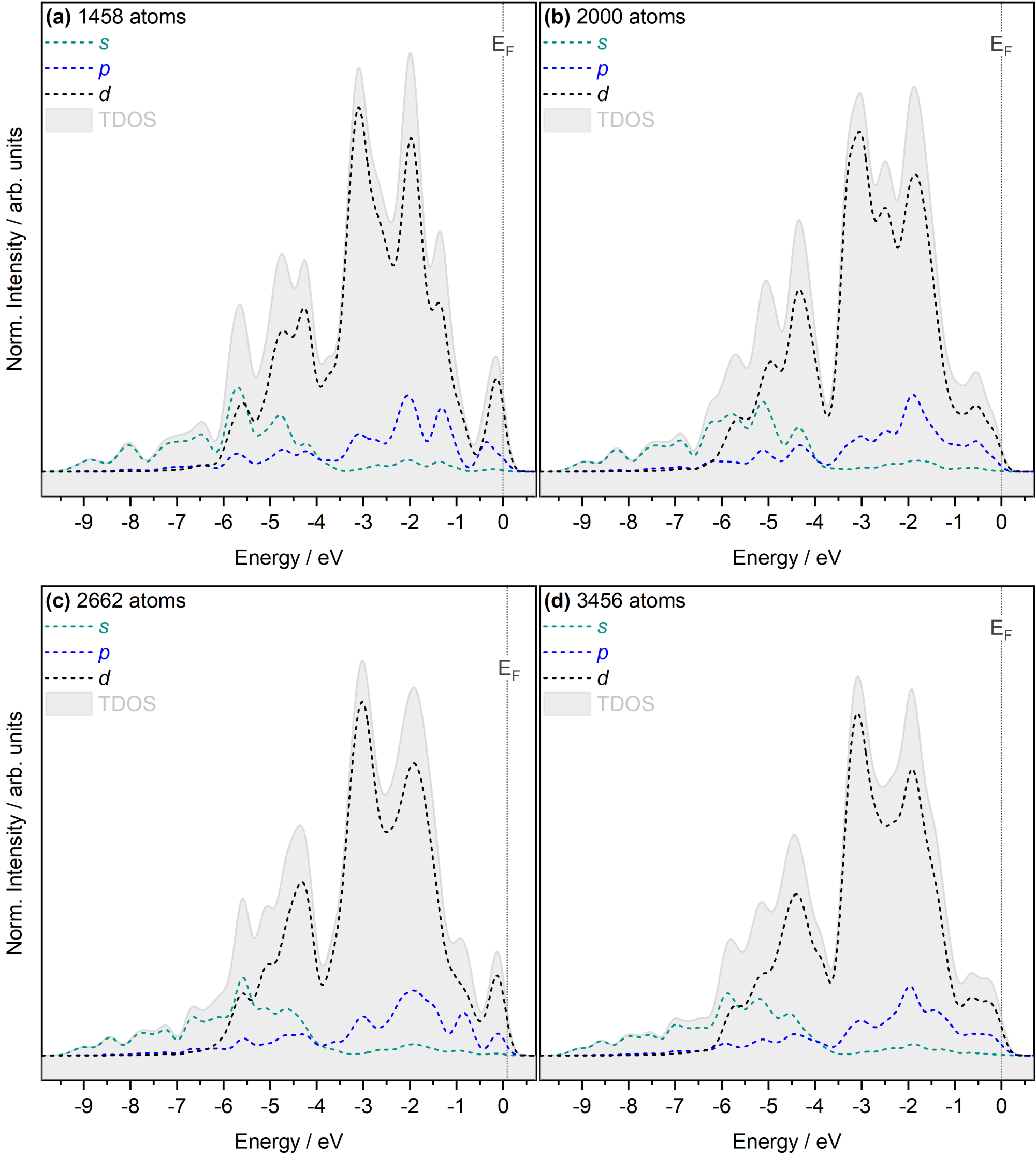}
    \caption{Unweighted PDOS calculated using DFT with a wavelet basis set. Four system sizes were tested to see the influence of supercell size: (a) 1458, (b) 2000, (c) 2662, and (d) 3456 atoms. TDOS refers to the sum of the PDOS.}
    \label{fig:Atom_Models}
\end{figure*}

\cleardoublepage

\section{Parameters Used to Construct Simulated Core Levels}\label{construction}
 The following Tables list the parameters used to construct the simulated core level spectra using the 4\textit{f} and/or 5\textit{p} spectral functions. The procedure of construction is as follows:\par
 
 \begin{enumerate}
  \item Apply Gaussian and Lorentzian broadening to the spectral function (for comparison to SXPS and HAXPES spectra, 350~meV and 266~meV Gaussian broadening was applied, respectively)
  \item Create a copy of the broadened spectral function
  \item Shift the copy according to the spin orbit splitting derived from HAXPES measurements
  \item Apply a scaling factor to the copy to ensure the degeneracy of the doublet is appropriate. This scaling factor is determined using the Scofield cross section tabulated data~\cite{Scofield1973TheoreticalKeV}
  \item Merge the copy and original spectral functions together to construct the spectrum
\end{enumerate}

The Lorentzian broadening of a core level is attributed to the natural line width generated due to the filling of a core hole. Campbell~\textit{et al.}\ have tabulated theoretical and experimental line widths of atomic K-N7 levels for tungsten and derived a recommended value based on the data available.~\cite{Campbell2001WidthsLevels} These recommended values were used to determine the level of Lorentzian broadening needed. The main theoretical dataset that Campbell \textit{et al.}\ compare the experimental data to is those calculated by Perkins \textit{et al.}\, known as EADL values.~\cite{Perkins1991Tables1-100} Campbell \textit{et al.}\ state that M shell data (i.e. 3\textit{s}, 3\textit{p}, 3\textit{d}) is predominately from XPS, but due to the inability of the M5 level (i.e. 3\textit{d}\textsubscript{5/2}) to de-excite by the Coster-Kronig transition, the theoretical EADL value calculated by Perkins \textit{et al.}\ is recommended. Furthermore, the EADL values were shown to be inappropriate for M4 levels (i.e. 3\textit{d}\textsubscript{3/2)} up to Z = 57 and instead the M5 EADL values should be taken as the M4 value. 5\textit{p} line widths were not reported by Campbell~\textit{et al.}\ and so the EADL values were chosen. Campbell~\textit{et al.}\ provide recommended values of the 4\textit{f} level, with the majority of the data derived from XPS measurements and are in good agreement with the EADL values. Tab.~\ref{lorentzian} lists the values used, along with the source of the data and the error of the value. Error estimates were reported by Campbell~\textit{et al.}\ for the recommended values.

\begin{table}[ht]
\caption{\label{lorentzian}Line widths of W core levels from Refs.~\cite{Perkins1991Tables1-100,Campbell2001WidthsLevels}}

    \begin{tabular}{clcc}
    \hline \hline
    $\textbf{Level}$ & $\textbf{Line~Width~/~eV}$ & $\textbf{Error Estimate}$ & $\textbf{Source}$ \\
    \hline
    3\textit{d}\textsubscript{5/2} & 1.7 & - & \cite{Perkins1991Tables1-100} \\
    3\textit{d}\textsubscript{3/2} & 1.7 & - & \cite{Campbell2001WidthsLevels} \\
    4\textit{d}\textsubscript{5/2} & 3.8 & $\pm$0.5~eV & \cite{Campbell2001WidthsLevels} \\
    4\textit{d}\textsubscript{3/2} & 4.1 & $\pm$0.5~eV & \cite{Campbell2001WidthsLevels} \\
    5\textit{p}\textsubscript{3/2} & 0.83 & - & \cite{Perkins1991Tables1-100} \\
    5\textit{p}\textsubscript{1/2} & 2.121 & - & \cite{Perkins1991Tables1-100} \\
    4\textit{f}\textsubscript{7/2} & 0.06 & $\pm$0.05~eV & \cite{Campbell2001WidthsLevels} \\
    4\textit{f}\textsubscript{5/2} & 0.1 & $\pm$0.05~eV & \cite{Campbell2001WidthsLevels} \\
    \hline \hline
    \end{tabular}

\end{table}

The scaling factors were determined from the Scofield tabulated data and the values used are listed in Tab.~\ref{Scaling}. For example, during the construction of the W~4\textit{f} and W~5\textit{p} simulated SXPS core level, the copy of the 4\textit{f} spectral function used to construct the W~4\textit{f}\textsubscript{5/2} core level was multiplied by 0.79, whereas the original 5\textit{p} spectral function used to construct the W~5\textit{p}\textsubscript{3/2} core level was multiplied by 0.15 to ensure the ratio between the W~4\textit{f}\textsubscript{7/2} and W~5\textit{p}\textsubscript{3/2} intensities were appropriate. These scaling factors are dependent on the photoionisation cross sections, which decay with increasing photon energy, and therefore, the scaling factor differs depending on the photon energy. Tab.~\ref{Scaling} lists the scaling factor that needs to be applied to the lower spin core level, as well as the scaling factor required for the W~5\textit{p}\textsubscript{3/2} relative to the W~4\textit{f}\textsubscript{7/2} core level, when using either Al-K$\alpha$ (SXPS) or 5.93~keV (HAXPES) photon energies.

\begin{table}[ht]
\caption{\label{Scaling}Scaling factors determined from the Scofield Cross Section Tabulated Data.~\cite{Scofield1973TheoreticalKeV}}

    \begin{tabular}{ccc}
    \hline \hline
    $\textbf{Level}$ & $\textbf{SXPS}$ & $\textbf{HAXPES}$ \\
    \hline
    3\textit{d} & - &  0.74 \\
    4\textit{d} & 0.69 & 0.75 \\
    5\textit{p} & 0.45 & 0.58 \\
    4\textit{f} & 0.79 & 0.81 \\
    5\textit{p}\textsubscript{3/2}/4\textit{f}\textsubscript{7/2} & 0.15 & 1.68 \\
    \hline \hline
    \end{tabular}
\end{table}

The spin orbit splitting distance to be applied was taken from the HAXPES experimental data and is listed in Tab.~\ref{SOS}.

\begin{table}[ht]
\caption{\label{SOS}Spin orbit splitting determined from HAXPES measurements.}

    \begin{tabular}{cr}
    \hline \hline
    $\textbf{Level}$ & $\textbf{Separation~/~eV}$ \\
    \hline
    3\textit{d} & 62.1  \\
    4\textit{d} & 12.5  \\
    5\textit{p} & 10.7  \\
    4\textit{f} & 2.2  \\
    5\textit{p}\textsubscript{3/2}-4\textit{f}\textsubscript{7/2} & 5.5  \\
    \hline \hline
    \end{tabular}

\end{table}

\cleardoublepage
\section{Further Photoionisation Cross Section Information}\label{Cross_Sections}

\begin{figure*}[ht!]
\centering
    \includegraphics[keepaspectratio, width=0.5\textwidth]{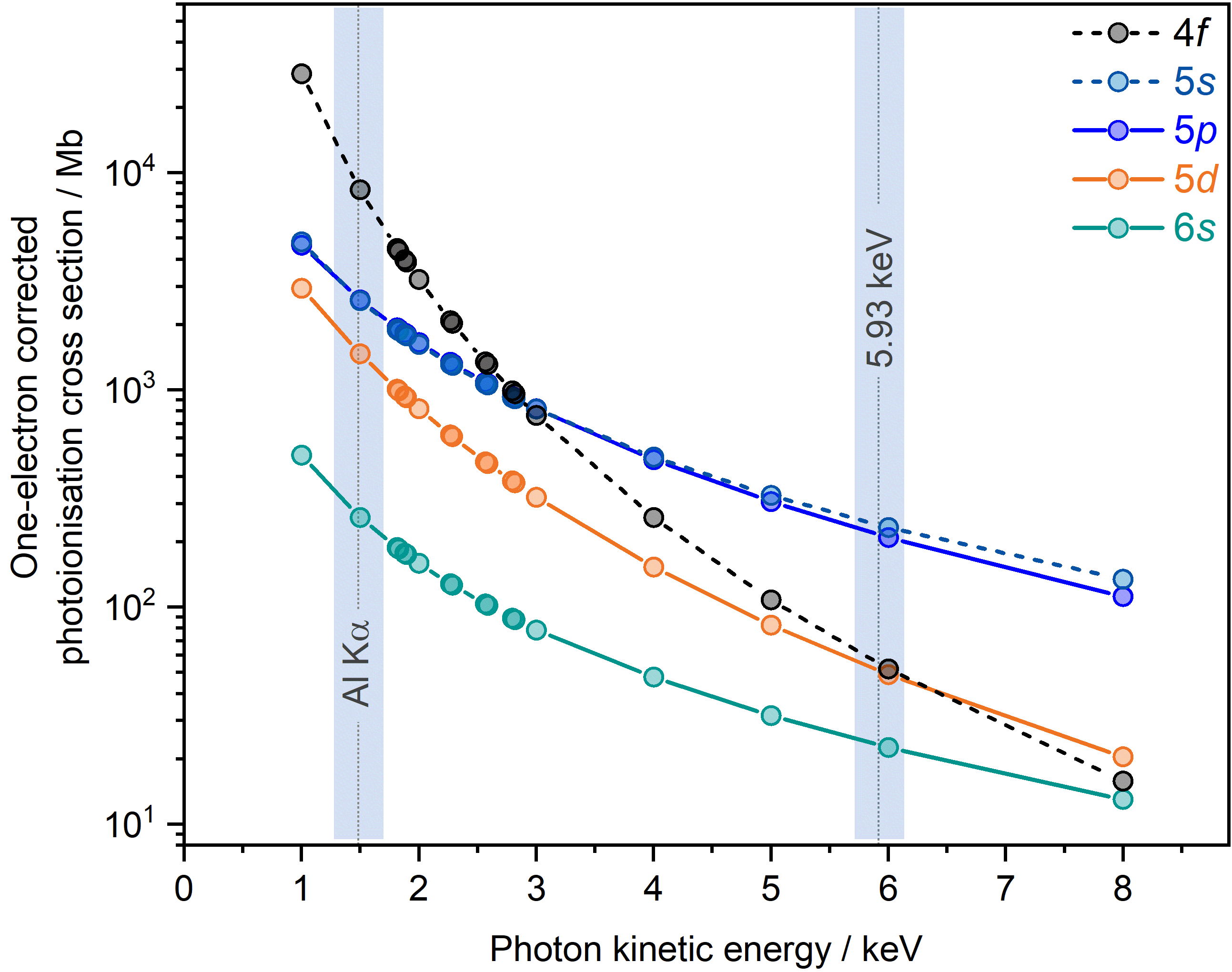}
    \caption{The one-electron corrected photoionisation cross sections of relevant atomic orbitals. It can be observed that at approximately 2.8~keV a cross-over between the~W 4\textit{f} and W~5\textit{p}/W~5\textit{s} is observed. The maximum kinetic energies (equal to the excitation energy) for Al K$_\alpha$ and 5.93 keV are highlighted.}
    \label{fig:X_section_W}
\end{figure*}

Fig.~\ref{fig:X_section_W} displays the one-electron photoionisation cross sections of key atomic orbitals of tungsten taken from Refs. \cite{Scofield1973TheoreticalKeV, Kalha20} as a function of photon energy. 

Tab.~\ref{X_Sections} lists the one-electron photoionisation cross sections values of interest when correcting the PDOS spectra for the comparison to the experimental valence band spectra. Tab.~\ref{X_Sections} also lists the correction factor determined when using the Pb core levels to estimate the W~6\textit{p} cross section (listed as ``Pb correction'') and the values obtained from the ``optimised'' method. The Pb correction cross section values are determined by multiplying the W~6\textit{s} cross section by the Pb~6\textit{p}/Pb~6\textit{s} ratio also listed in the table. The optimisation method is sensitive to both the projection and the smoothness of the PDOS. Therefore, due to the similarity between their PDOS, the optimum \textit{p} cross section determined by optimising the plane-wave DFT PDOS was very similar to the G0W0 value, while the wavelet DFT value differed significantly. Since the G0W0 PDOS is considerably smoother than the wavelet DFT PDOS, due to the larger effective $k$-point sampling, the ``optimised'' value listed in Table~\ref{X_Sections} is taken from G0W0.

\begin{table}[ht!]
    \caption{\label{X_Sections}Tabulated one-electron photoionisation cross sections $\sigma_i$ of interest for W and Pb taken from Refs.~\cite{Scofield1973TheoreticalKeV, Kalha20}.}
    \begin{tabular}{ccc}
    \hline \hline
    Orbital & $\sigma_i$ for Al K{$\alpha$} / Mb/e\textsuperscript{-} & $\sigma_i$ for 5.9~keV / Mb/e\textsuperscript{-}\\
    \hline
    W~5\textit{p} & 2.633e+3 & 2.146e+2\\
    W~6\textit{s} & 2.620e+2 & 2.293e+1\\
    W~5\textit{d} & 1.494e+3 & 5.063e+1\\
    Pb~6\textit{s} & 5.121e+2 & 4.695e+1\\
    Pb~6\textit{p} & 2.959e+2 & 2.804e+1\\
    \hline
    Pb~6\textit{p}/Pb~6\textit{s} ratio & 0.579 & 0.607\\
    Pb correction & 1.52e+2 & 1.39e+1\\
    \hline
    Optimised & 3.449e+3 & 1.449e+2\\
    \hline \hline
    \end{tabular}
\end{table}

\begin{figure*}[ht!]
\centering
    \includegraphics[keepaspectratio, width=\textwidth]{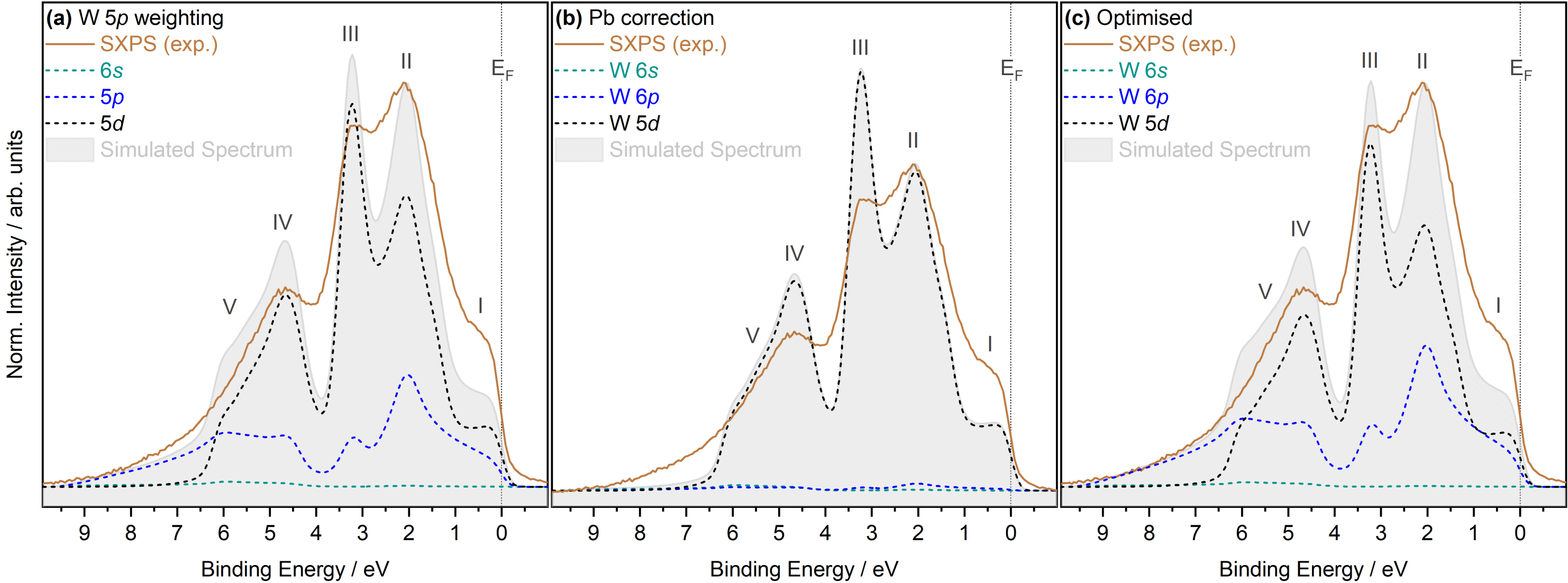}
    \caption{Comparison of the three methods used to apply photoionisation cross sections to the calculated PDOS. The plots compare the PDOS calculated using DFT with a plane-wave basis set to the experimental valence band collected with SXPS. The PDOS was suitably broadened to match the experimental broadening, whereas a Shirley-type function was applied to the experimental spectra to remove the background. The spectra are normalised to the intensity of feature (II). Simulated Spectrum refers to the sum of the cross-section weighted PDOS.}
    \label{fig:Xsection}
\end{figure*}

Fig.~\ref{fig:Xsection} displays the comparison of all three photoionisation cross section correction methods applied to the PDOS calculated using the plane-wave DFT code. The 5\textit{p} correction approach shown in Fig.~\ref{fig:Xsection}(a) appears suitable but physically difficult to justify. The Pb approach shown in Fig.~\ref{fig:Xsection}(b) clearly under predicts the weighting, allowing the \textit{d} state to completely dominate the spectra. The final, optimised fitting approach (shown in Fig.~\ref{fig:Xsection}(c)) is further discussed in the main manuscript.

\cleardoublepage

\section{Survey Spectrum}\label{Surv}

\cleardoublepage

\section{SXPS and HAXPES Core Level Details}\label{BE_FWHM}

\begin{table}[ht!]
\caption{\label{W_BE}Absolute binding energy (BE) positions of tungsten core level and satellite peak positions, spin orbit splitting (SOS) determined from HAXPES, and full width at half maximum (FWHM), determined from SXPS and HAXPES (core level BE position and FWHM error = $\pm$0.05~eV, satellite BE position error = $\pm$0.1~eV)}
\begin{tabular}{cccccc}
\hline \hline
$\textbf{Peak}$ & $\textbf{BE~Position~ SXPS~/~eV}$ & $\textbf{BE~Position~ HAXPES~/~eV}$ & $\textbf{Spin~Orbit~Splitting~/~eV}$ & $\textbf{FWHM~SXPS~/~eV}$ & $\textbf{FWHM~HAXPES~/~eV}$\\
\hline
W 4\textit{f}\textsubscript{7/2} & 31.4 & 31.3 & - & 0.5 & 0.4\\
W 4\textit{f}\textsubscript{5/2} & 33.6 & 33.5 & 2.2 & 0.5& 0.4\\
W 5\textit{p}\textsubscript{3/2} & 36.9 & 36.9 & - & -& 1.7\\
S\textsubscript{1} & 42.7 & 42.7 & - & - & -\\
W 5\textit{p}\textsubscript{1/2} & 47.3 & 47.6 & 10.7 & - & 4.2 \\
S\textsubscript{2} & 57.3 & 57.3 & - & - & -\\
S\textsubscript{3} & - & 63.1 & - & - & -\\
S\textsubscript{4} & 66.1 & 66.1 & - & -& -\\
W 5\textit{s} & 75.4 & 75.2 & - & - & 4.6\\

W 4\textit{d}\textsubscript{5/2} & 243.5 & 243.4 & - & 4.6 & 4.0\\
W 4\textit{d}\textsubscript{3/2} & 256.0 & 255.9 & 12.5 & 5.3 & 4.5 \\
S\textsubscript{1} & 268.1 & 268.1 & - & -& -\\
S\textsubscript{2} & 281.7 & 281.7 & - & - & -\\
S\textsubscript{3} & 297.1 & 297.1 & - & - & -\\

W 3\textit{d}\textsubscript{5/2} & - & 1806.3 & - & - &3.4 \\
S\textsubscript{1} & - & 1817.8 & - & - & - \\
S\textsubscript{2} & - & 1831.5 & - & - & -\\
S\textsubscript{3} & - & 1852.4 & - & - & -\\
W 3\textit{d}\textsubscript{3/2} & - & 1868.4 & 62.1 & - & 4.6 \\
S\textsubscript{4} & - & 1879.8 & - & - & -\\
S\textsubscript{5} & - & 1893.1 & - & - & -\\
S\textsubscript{6} & - & 1919.9 & - & - & -\\
W 3\textit{p}\textsubscript{3/2} & - & 2277.5 & - & - & 10.8\\
S\textsubscript{1} & - & 2302.1 & - & - & -\\
S\textsubscript{2} & - & 2328.8 & - & - & -\\
W 3\textit{p}\textsubscript{1/2} & - & 2571.3 & 293.8 & - & 14.5\\
S\textsubscript{1} & - & 2595.9 & - & - & -\\
    
W 3\textit{s} & - & 2817.3 & - & - & 16.5 \\
S\textsubscript{1} & - & 2841.7 & - & - &-\\
\hline \hline

\end{tabular}
\end{table}

\cleardoublepage

\section{G0W0 and GW+C Calculated Unweighted PDOS}\label{Unweighted_PDOS_GW}

\begin{figure*}[ht!]
\centering
    \includegraphics[keepaspectratio, width=0.6\textwidth]{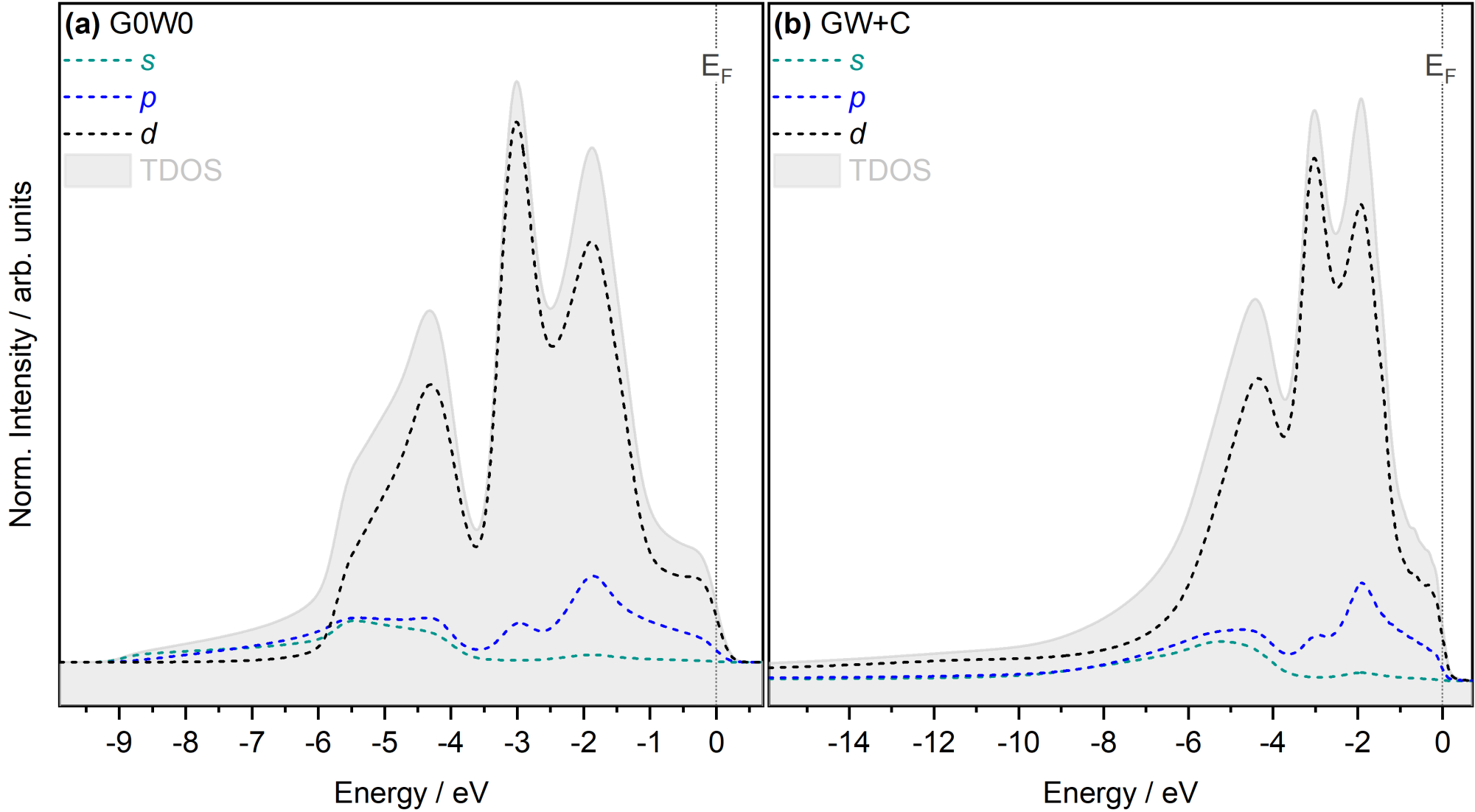}
    \caption{Unweighted PDOS spectra calculated using the (a) G0W0 and (b) GW+C approaches. TDOS refers to the sum of the PDOS.}
    \label{fig:unweighted_PDOS}
\end{figure*}

\cleardoublepage

\section{Comparison of PDOS to SXPS Valence Band}\label{SXPS_PDOS}

\begin{figure*}[ht!]
\centering
    \includegraphics[keepaspectratio, width=\textwidth]{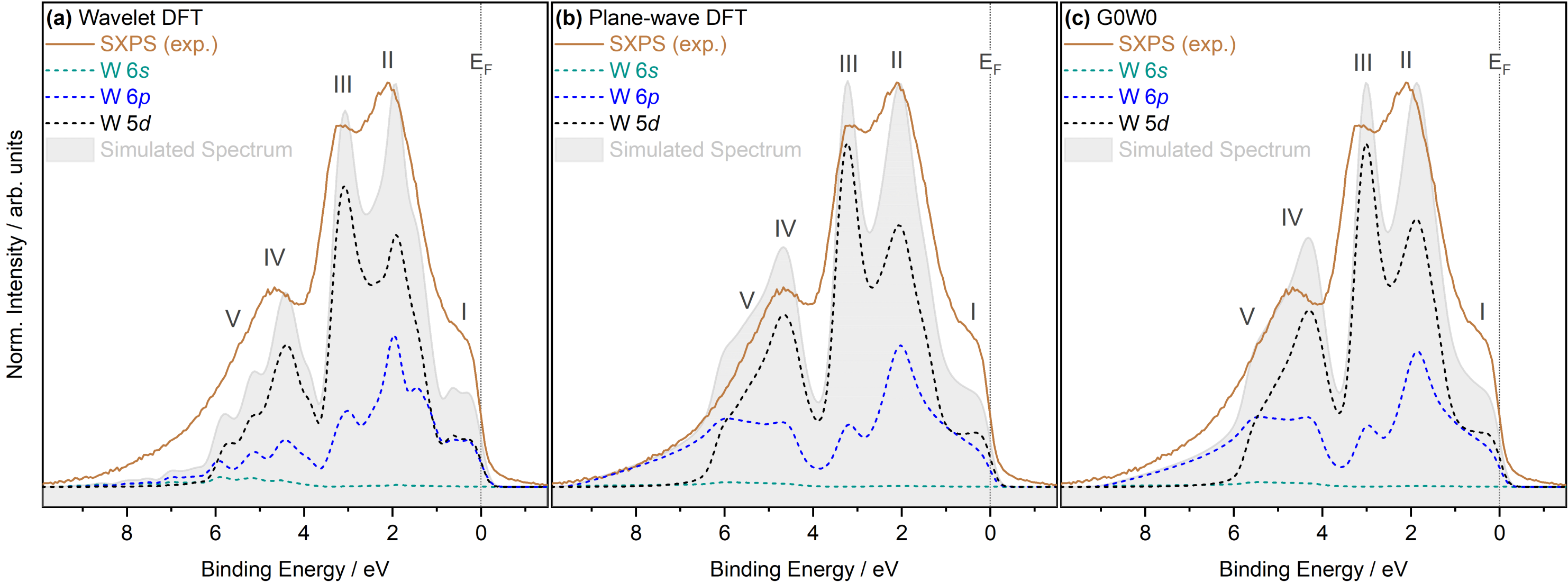}
    \caption{Comparison of simulated PDOS spectra calculated using DFT and G0W0 approaches with the SXPS valence band spectra, including the comparison with (a) DFT using a wavelet basis set, (b) DFT using a plane-wave basis set, and (c) G0W0. A Shirley-type background was removed from the experimental spectrum to allow direct comparison to the theory. The PDOS contributions have been cross section weighted and suitably broadened to match the experimental broadening. Simulated Spectrum refers to the sum of the PDOS.}
    \label{fig:SX_PDOS}
\end{figure*}

\section{Comparison of DFT Methods}\label{DFT_compare}

\begin{figure*}[ht]
\centering
    \includegraphics[keepaspectratio, width = 0.95\linewidth]{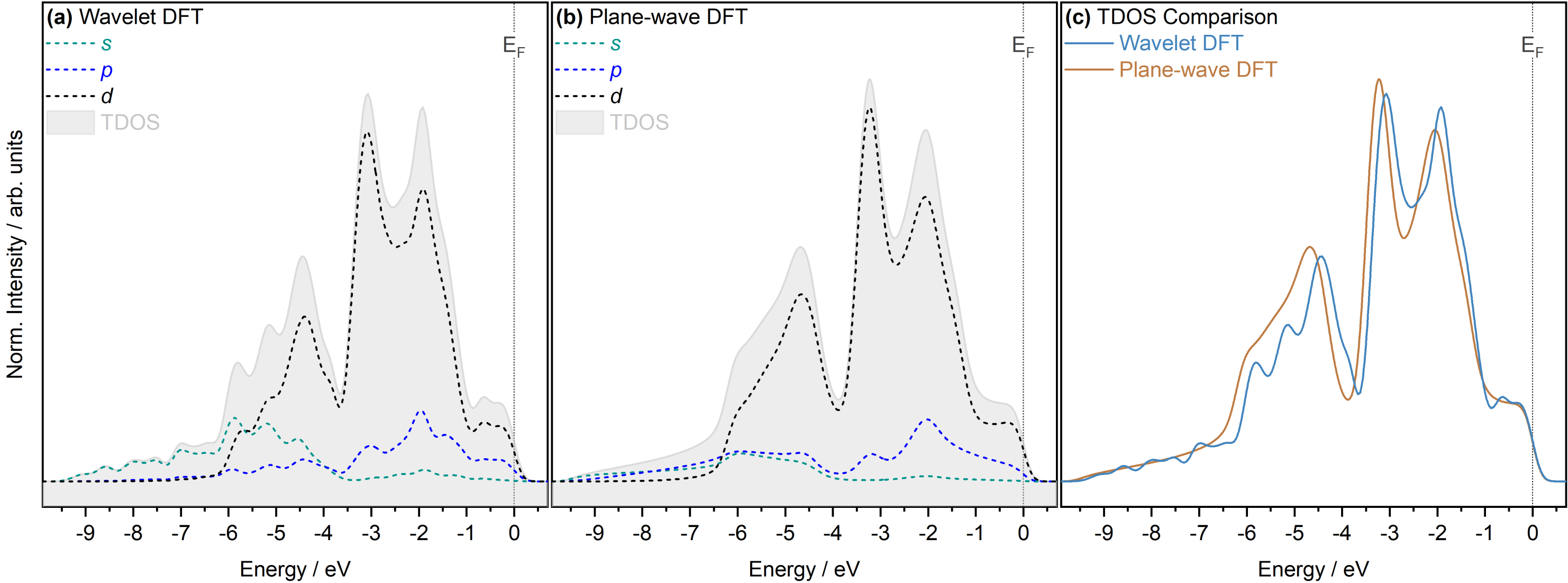}
    \caption{Unweighted PDOS calculated from both DFT approaches. (a) PDOS calculated from a 3456 atom supercell with a 1$\times$1$\times$1 \textit{k}-point grid using DFT that employed a wavelet basis, (b) PDOS calculated using DFT that employed a plane-wave basis with a \textit{k}-point 64$\times$64$\times$64 grid and primitive unit cell, and (c) a direct comparison of the total density of states (TDOS) from both approaches. The PDOS are aligned to the calculated Fermi edge. The PDOS shown in (c) are normalised with respect to their total areas to allow for a direct comparison. TDOS refers to the sum of the PDOS. Unoccupied states are excluded and the PDOS shown are not photoionisation cross section corrected. Additionally, the data is broadened with a full width at half maximum (FWHM) of 0.35~eV using a Gaussian function.}
    \label{fig:DFT_Methods}
\end{figure*}

\cleardoublepage

\bibliography{references_SI.bib}
\bibliographystyle{apsrev4-1}